\pdfoutput=1
\documentclass[aps,prd,amsmath,floats,floatfix, twocolumn,
superscriptaddress,nofootinbib,showpacs,longbibliography]{revtex4-1}
\usepackage[T1]{fontenc}
\usepackage[utf8]{inputenc}
\usepackage{lmodern}

\usepackage{verbatim}

\usepackage[dvipsnames, usenames]{xcolor}
\definecolor{linkcolor}{rgb}{0.0,0.3,0.5}
\usepackage[hypertexnames=false, unicode, colorlinks=true, linkcolor=linkcolor,
citecolor=linkcolor, filecolor=linkcolor,urlcolor=linkcolor,
pdfusetitle]{hyperref}

\usepackage[all]{hypcap}
\usepackage{graphicx}
\usepackage{xspace}
\usepackage{amssymb}
\usepackage[normalem]{ulem}
\usepackage{bm}

\usepackage{microtype}

\usepackage[english]{babel}
\usepackage{blindtext}

\newcommand\prlsec[1]{\vspace{2mm}\noindent \emph{#1}.---}

\graphicspath{%
  {figs/}%
}

\DeclareMathAlphabet{\mathpzc}{OT1}{pzc}{m}{it}

\newcommand{\roughly}{\mathchar"5218\relax\,}

\newcommand{\bchi}{\bm{\chi}}
\newcommand{\bv}{\bm{v}}
\newcommand{\bL}{\bm{\Lambda}}

\newcommand{\NRSur}{\texttt{NRSur7dq4}\xspace}
\newcommand{\NRSurRemnant}{\texttt{NRSur7dq4Remnant}\xspace}

\newcommand{\sinekicksfignum}{3\xspace}
\newcommand{\violinfignum}{4\xspace}
\newcommand{\dopplerfignum}{6\xspace}

\newcommand{\TITLE}{Extracting the Gravitational Recoil from Black Hole Merger
Signals}

\newcommand\caltech{\affiliation{TAPIR 350-17, California Institute of
Technology, 1200 E California Boulevard, Pasadena, CA 91125, USA}}
\newcommand\MIT{\affiliation{LIGO Laboratory, Massachusetts Institute of
Technology, Cambridge, Massachusetts 02139, USA}}

\newcommand{\dcc}{LIGO-P2000030}

\begin{document}

\title{\TITLE}

\author{Vijay Varma}
\email{vvarma@caltech.edu}
\caltech

\author{Maximiliano Isi}
\email[]{maxisi@mit.edu}
\thanks{NHFP Einstein fellow}
\MIT

\author{Sylvia Biscoveanu}
\email{sbisco@mit.edu}
\MIT

\hypersetup{pdfauthor={Varma, Isi, and Biscoveanu}}

\date{\today}

\begin{abstract}
Gravitational waves carry energy, angular momentum, and linear momentum.  In
generic binary black hole mergers, the loss of linear momentum imparts a recoil
velocity, or a ``kick'', to the remnant black hole. We exploit recent advances
in gravitational waveform and remnant black hole modeling to extract
information about the kick from the gravitational wave signal. Kick
measurements such as these are astrophysically valuable, enabling independent
constraints on the rate of second-generation mergers. Further, we show that
kicks must be factored into future ringdown tests of general relativity with
third-generation gravitational wave detectors to avoid systematic biases. We
find that, although little information can be gained about the kick for
existing gravitational wave events, interesting measurements will soon become
possible as detectors improve. We show that, once LIGO and Virgo reach their
design sensitivities, we will reliably extract the kick velocity for
generically precessing binaries---including the so-called superkicks, reaching
up to 5000 km/s.
\end{abstract}

\maketitle

\prlsec{Introduction}
As existing gravitational wave (GW) detectors, Advanced
LIGO~\cite{TheLIGOScientific:2014jea} and Virgo~\cite{TheVirgo:2014hva},
approach their design sensitivities, they continue to open up unprecedented
avenues for studying the astrophysics of black holes (BHs). One such
opportunity is to experimentally study the gravitational recoil in binary BH
mergers. It is well known that GWs carry away energy and angular momentum,
causing the binary to shrink during the inspiral; however, in addition to this,
GWs also carry away linear momentum, shifting the binary's center of mass in
the opposite direction~\cite{Bonnor:1961linmom, PhysRev.128.2471,
Bekenstein:1973ApJ, Fitchett:1983MNRAS}. Learning about this effect from GW
data would be of high astrophysical significance.

During a binary BH coalescence, most of the linear momentum is radiated near
the time of the merger~\cite{Gonzalez:2006md, Lousto:2007db, Lousto:2012su,
Lousto:2012gt, Blanchet:2005rj, Damour:2006tr, LeTiec:2009yg}, resulting in a
recoil or ``kick'' imparted to the remnant BH.  The end state of the remnant is
entirely characterized by its mass ($m_f$), spin ($\bchi_f$) and kick velocity
($\bv_f$); all additional complexities (``hair'')~\cite{1968_Israel,
PhysRevLett.26.331} are dissipated away in GWs during the ringdown stage that
follows the merger. The remnant mass and spin have already been measured from
GW signals and used to test general relativity~\cite{TheLIGOScientific:2016src,
    LIGOScientific:2019fpa, Ghosh:2017gfp, Brito:2018rfr, Carullo:2018sfu,
Carullo:2019flw, Isi:2019aib, Giesler:2019uxc}. However, a measurement of the
kick has remained elusive.

Measuring the kick velocity from binary BHs would have important astrophysical
applications---particularly for precessing binaries, where the component BH
spins have generic orientations with respect to the orbit. For these systems,
the spins interact with the orbital angular momentum as well as with each
other, causing the orbital plane to precess~\cite{Apostolatos:1994pre}.  The
kick velocity of these systems can reach up to $5000$ km/s for certain
fine-tuned configurations~\cite{Campanelli:2007cga, Gonzalez:2007hi,
Tichy:2007hk, Lousto:2011kp, Lousto:2019lyf, Sperhake:2019wwo}, earning them
the moniker of ``superkicks''.  Such velocities are larger than the escape
velocity of even the most massive galaxies. This can have dramatic consequences
for mergers of supermassive BHs residing at galactic centers.  The remnant BH
can be significantly displaced or ejected~\cite{Merritt:2004xa,
Komossa:2008qd}, impacting the galaxy's evolution~\cite{Volonteri:2010mbh,
Komossa:2008as, Gerosa:2014gja}, and event rates~\cite{sesana:2007zk} for the
future LISA mission~\cite{amaroseoane2017laser}.

The kick velocity is also important for second-generation stellar-mass mergers,
where one of the component BHs originated in a previous merger. This scenario
has attracted much attention recently~\cite{Yang:2019cbr, Gayathri:2019kop,
    Gerosa:2019zmo, Mangiagli:2019sxg, DiCarlo:2019fcq, Rodriguez:2019huv,
Doctor:2019ruh, Farmer:2019jed} because the GW event
GW170729~\cite{LIGOScientific:2018mvr, Chatziioannou:2019dsz} may have a
component BH that is too massive to originate in a supernova
explosion~\cite{Woosley:2016hmi, Marchant:2018kun}, the typical formation
scenario for stellar-mass BHs. A second-generation merger could explain this,
as the first merger would have led to a remnant BH more massive than the
original stellar-mass progenitors. If we could measure the kick velocity from
GW signals, we could place independent constraints on rates of
second-generation mergers.

In this \emph{Letter}, we present the first method to extract the kick
magnitude and direction from generically precessing GW signals. We demonstrate
that kicks will be measured reliably once LIGO and Virgo reach their design
sensitivities, and possibly even earlier. The key is being able to accurately
measure the spins of the individual BHs in the binary, from which the kick
velocity can be inferred.  This is made possible by two advances in GW modeling
achieved in the past few years: numerical relativity (NR) surrogate models for
both gravitational waveforms~\cite{Varma:2019csw, Blackman:2017pcm} and remnant
properties~\cite{Varma:2019csw, Varma:2018aht}, suitable for generically
precessing binaries. These models capture the effects of spin precession at an
accuracy level comparable to the NR simulations, and are the most accurate
models currently available in their regime of validity~\cite{Varma:2019csw}.

\prlsec{Methods} We use the surrogate waveform model
\NRSur~\cite{Varma:2019csw} to analyze public GW
data~\cite{LIGOScientific:2018mvr, GWOSC:GWTC}, as well as simulated signals in
synthetic Gaussian noise corresponding to the three-detector advanced
LIGO-Virgo network at its design
sensitivity~\cite{aLIGODesignNoiseCurve,Manzotti:2012uw,Aasi:2013wya}.

\NRSur is trained on NR simulations with mass ratios $q=m_1/m_2 \leq 4$ and
component spin magnitudes $|\bchi_1|,|\bchi_2|\leq0.8$ with generic spin
directions.  The index 1 (2) corresponds to the heavier (lighter) BH, with $m_1
\geq m_2$.  The spin components are specified at a reference GW frequency
$f_{\text{ref}}=20$ Hz, in a source frame defined as follows: the $z$-axis lies
along the instantaneous orbital angular momentum, the $x$-axis points from the
lighter to the heavier BH, and the $y$-axis completes the right-handed triad.
We use all available spin-weighted spherical harmonic modes for this model
($\ell\leq4$). The inclination angle $\iota$ and azimuthal angle
$\phi_{\text{ref}}$ indicate the location of the observer in the sky of the
source, and take different values for each injection.

We obtain Bayesian posteriors on the signal parameters using the
\textsc{LALInference} package \cite{Veitch:2014wba}, part of the LIGO Algorithm
Library (LAL) Suite \cite{lalsuite}. Because of restrictions on the duration of
\NRSur waveforms, we choose to analyze data with a minimum Fourier frequency
$f_{\rm low} = 20$ Hz. Waveform length also restricts the higher-order-mode
content of our \NRSur injections and templates in such way that modes with
azimuthal harmonic number $m$ contribute with a starting frequency
$f^{(m)}_{\rm min} = m f_{\rm low}/2$. This means that our sensitivity
projections are conservative, as detectors are expected to access information
starting at lower frequencies than our simulations.  NR injections are handled
via the dedicated infrastructure in LAL~\cite{Schmidt:2017btt}.

Given the posteriors distributions for the component parameters $\bL = \{m_1,
m_2, \bchi_1, \bchi_2\}$, we use the remnant-properties surrogate model
\NRSurRemnant~\cite{Varma:2019csw} to predict the mass $m_f$, spin $\bchi_f$,
and kick velocity $\bv_f$ of the remnant. Trained on the same simulations as
\NRSur, \NRSurRemnant uses Gaussian Process
Regression~\cite{Rasmussen_Williams_GPRbook, Varma:2018aht} to model the
remnant properties. \NRSurRemnant improves upon previous remnant properties
models by at least an order of magnitude in accuracy~\cite{Varma:2019csw}.
\NRSurRemnant models the full kick velocity vector and can, therefore, predict
both the kick magnitude and direction. To assess whether a meaningful kick
measurement has been made, we compare this posterior distribution with the
corresponding \emph{effective} prior distribution, estimated by drawing
component parameters $\bL$ from the prior. The priors used for the component
parameters are discribed in the
Supplement~\cite{kickpapersupplement}.
\nocite{Krolak1987, Barausse:2012qz, Hofmann:2016yih, Jimenez-Forteza:2016oae,
Healy:2016lce, Healy:2014yta, Vishveshwara:1970, Press:1971ApJ,
Teukolsky:1973ApJ, Chandrasekhar_Detweiler:1975, Gair:2015nga, Dreyer:2003bv,
Gossan:2011ha, Meidam:2014jpa}

\prlsec{Comparison to previous methods}
The challenge of measuring the kick velocity from GW signals has been tackled
before. The recoil may Doppler shift the final portion of the GW signal.
Ref.~\cite{Gerosa:2016vip} showed that it will not be possible to measure the
kick velocity from this effect alone until third-generation GW detectors become
active in the 2030s~\cite{Reitze:2019iox, Punturo:2010zz,
Punturo:2010science_reach, Evans:2016mbw}. Ref.~\cite{CalderonBustillo:2018zuq}
proposed a method to extract the kick based on direct comparison against NR
simulations, showing that current detectors are sufficient for a kick
measurement; however, that study was restricted to nonprecessing systems, where
we do not expect very large kicks ($\gtrsim 300$ km/s).  Ref.~\cite{Healy:2019}
compared GW150914 data against NR simulations, including precessing systems, to
place bounds on the kick of GW150914. However, both
Refs.~\cite{CalderonBustillo:2018zuq} and \cite{Healy:2019} relied on a
discrete bank of NR simulations, which does not allow for a full exploration of
the multidimensional posterior for the system parameters.

\begin{figure}[tb]
\includegraphics[width=0.45\textwidth]{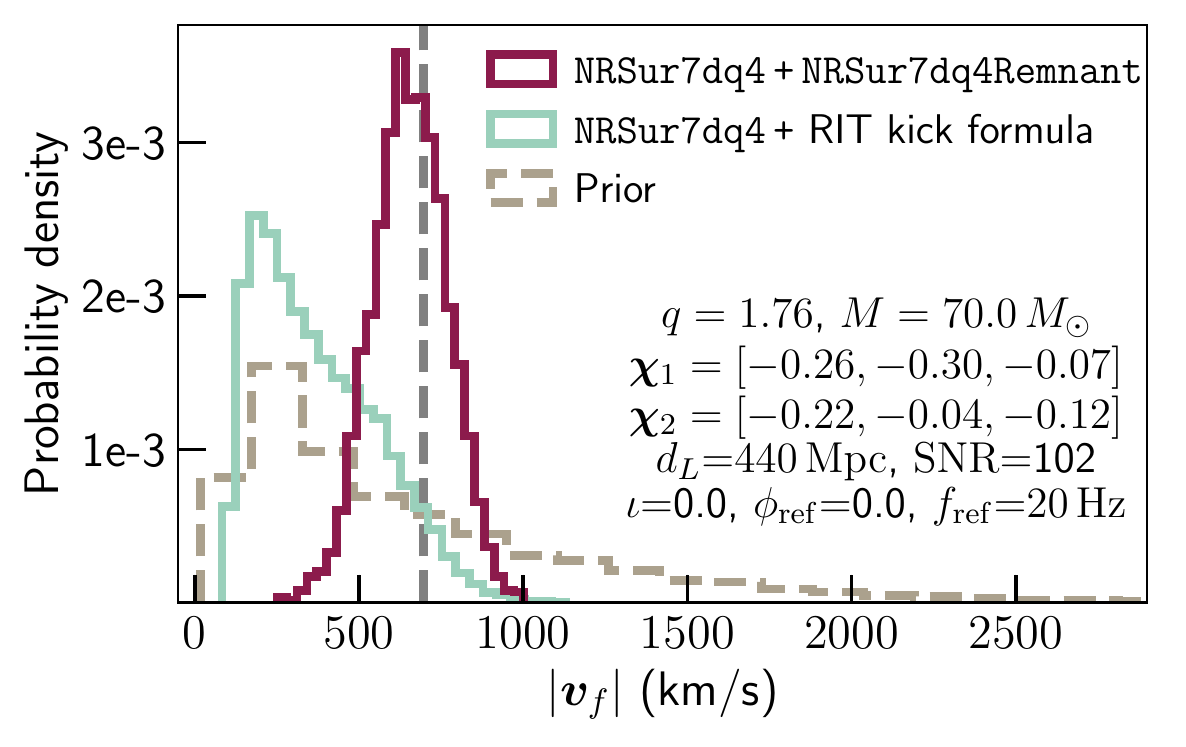}
\caption{Kick magnitude measurement using different remnant BH models in
    conjunction with the \NRSur waveform model, for an injected NR signal at
    the design sensitivity of LIGO and Virgo. The signal parameters are given
    in the inset text and the corresponding kick magnitude is indicated by the
    dashed gray line. The effective prior is shown as a dashed histogram.
}
\label{fig:compare_models}
\end{figure}

\begin{figure}[tb]
\includegraphics[width=0.42\textwidth]{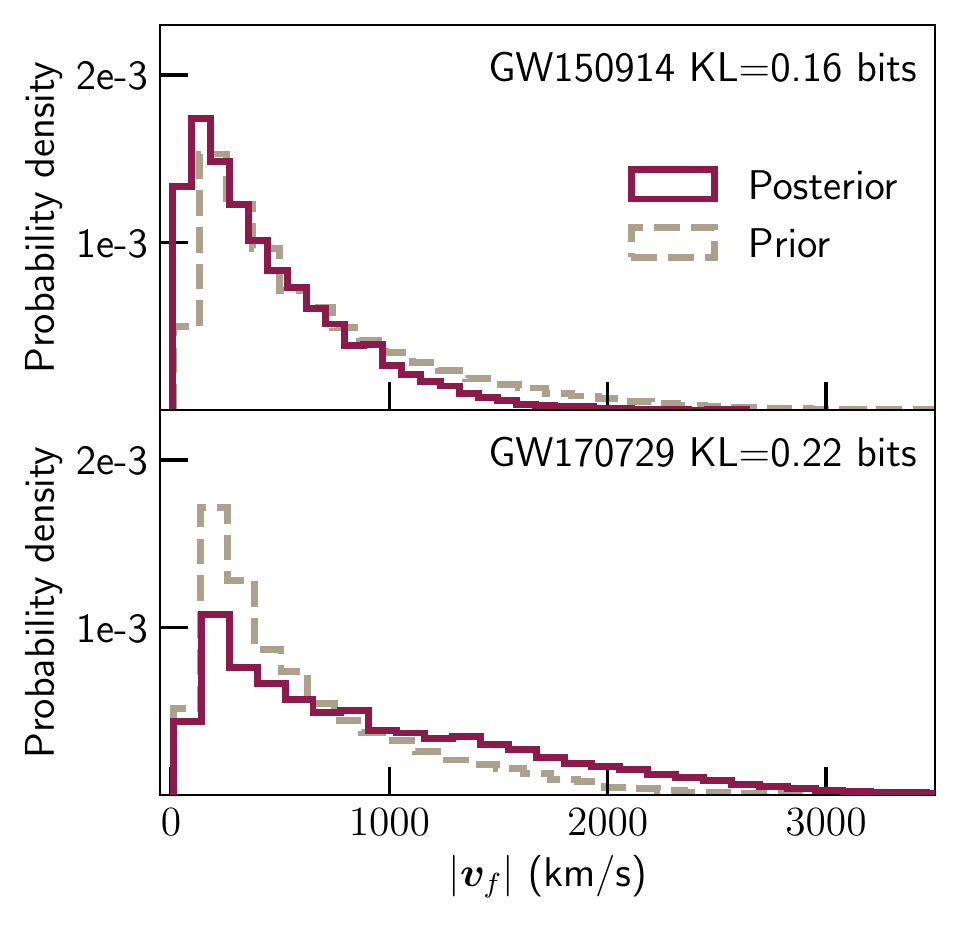}
\caption{Kick measurement for the GW events GW150914 and GW170729. We find only
    marginal differences between the posterior and the effective prior,
    suggesting that very little information about the kick can be gained from
    these events. We quantify this via the KL divergence, shown in the
    upper-right insets.
}
\label{fig:GWTC_events}
\end{figure}

\begin{figure*}[thb]
\includegraphics[width=0.45\textwidth]{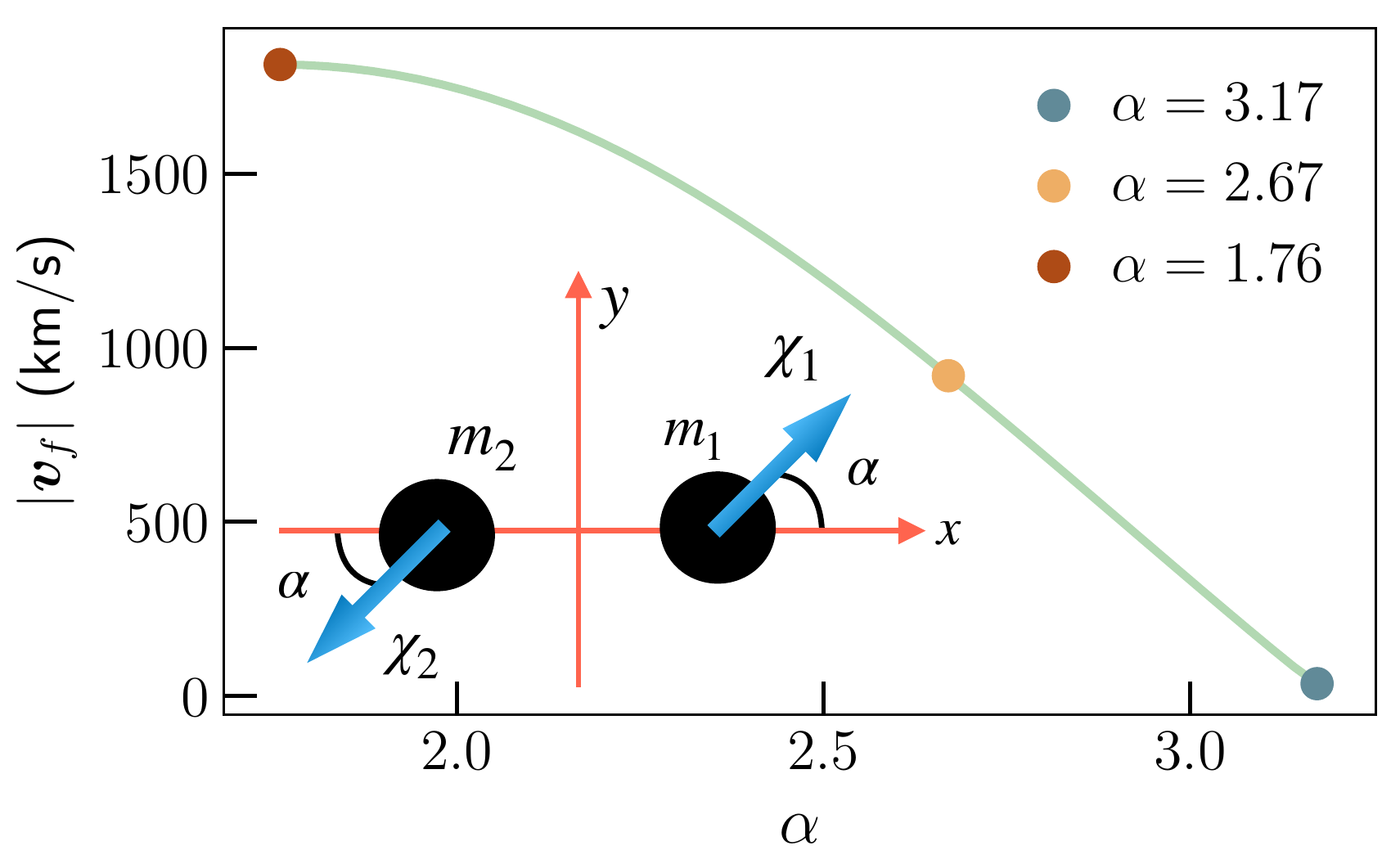}
\hspace{0.7cm}\includegraphics[width=0.432\textwidth]{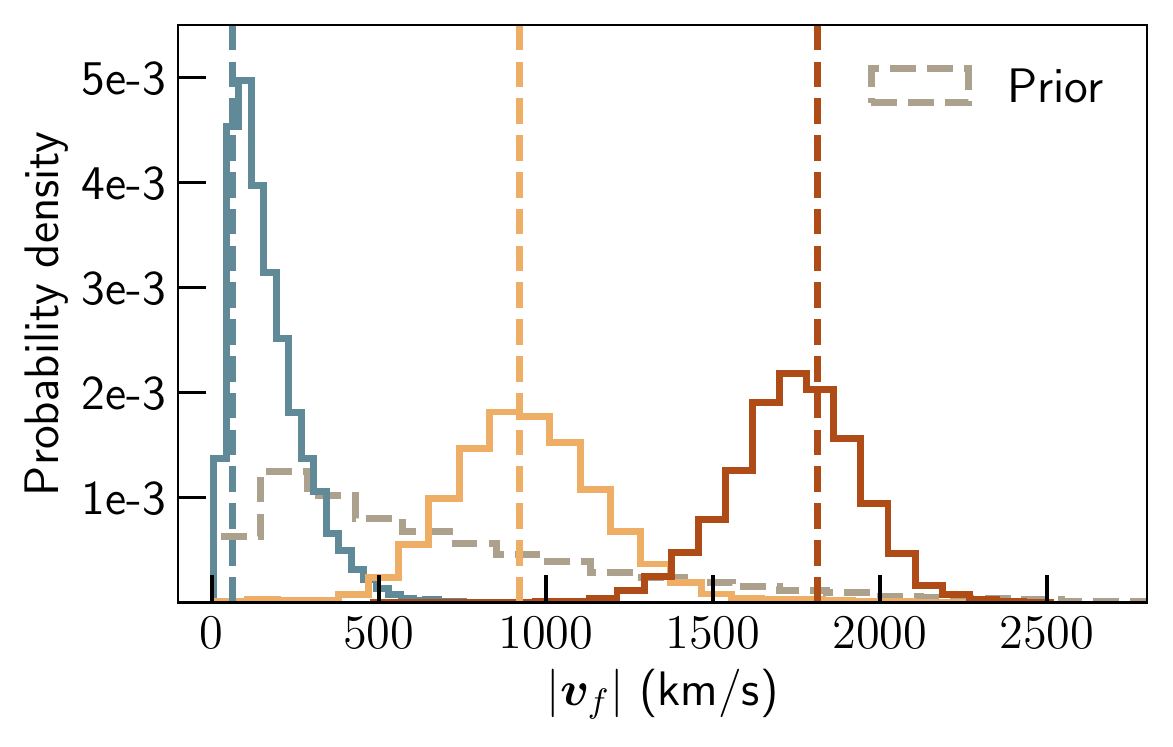}
\caption{A demonstration of the measurability of superkicks at the design
    sensitivity of LIGO and Virgo. We consider the fine-tuned binary
    configuration shown in the inset of the left panel. The kick velocity has a
    sinusoidal dependence on the angle $\alpha$ as shown in the left panel. We
    inject \NRSur signals corresponding to the three markers and measure the
    kick velocity using our method. The posteriors for the measured kick
    magnitudes are shown in the right panel; the colors correspond to the
    markers in the left panel. The true kick magnitudes are shown as dashed
    vertical lines, and the effective prior is shown as a dashed histogram. In
    all three cases, the kick velocity is well recovered.
}
\label{fig:superkick}
\end{figure*}

Our procedure for measuring kicks is more widely applicable than those of
Refs.~\cite{Gerosa:2016vip, CalderonBustillo:2018zuq, Healy:2019} in a few
ways. Since the surrogate models accurately reproduce the NR simulations, we
are potentially sensitive to effects of the recoil other than simple Doppler
shifts (e.g. acceleration of the center of mass near
merger~\cite{Gonzalez:2006md, Lousto:2007db, Lousto:2012su, Lousto:2012gt,
Blanchet:2005rj, Damour:2006tr, LeTiec:2009yg}, or phase aberration
\cite{Torres-Orjuela:2020cly}.). Therefore, rather than rely on Doppler shifts
in the ringdown~\cite{Gerosa:2016vip}, we instead extract information from the
full waveform. Based on the inferred binary parameters $\bL$, we infer the kick
using the \NRSurRemnant model.  \NRSurRemnant can take as input $\bL$
posteriors obtained with any waveform or inference setup. This allows us to
fully sample the posterior space, which cannot be covered by discrete NR
template banks~\cite{CalderonBustillo:2018zuq, Healy:2019}. Critically, our
method applies to precessing binaries where large kicks occur. As demonstrated
in the following sections, our method will soon make it possible to extract
kicks from generically precessing systems, including superkicks, in a fully
Bayesian setup.

\prlsec{NR simulation}
We first demonstrate our method by injecting an NR waveform into noise from a
simulated LIGO-Virgo network at design sensitivity. The signal parameters
are given in the inset text of Fig.~\ref{fig:compare_models}. We choose a
luminosity distance consistent with that of GW150914~\cite{Abbott:2016blz},
$d_L=440$ Mpc. Using the \NRSur waveform model, we recover the signal with a
signal-to-noise ratio (SNR) of 102. Here and throughout this paper, reported
SNRs correspond to the network matched-filter, maximum \emph{a-posteriori}
values. Further, all masses are reported in the detector frame.

Our method successfully recovers the injected kick magnitude, as seen from the
posterior in Fig.~\ref{fig:compare_models}. We find that the use of the remnant
surrogate model \NRSurRemnant is critical. To show this, we consider an
alternate kick formula developed in Refs.~\cite{Gonzalez:2006md,
Campanelli:2007ew, Lousto:2007db, Lousto:2012su, Lousto:2012gt}, as summarized
in \cite{Gerosa:2016sys}. Using this formula (which we label ``RIT'') on the
same \NRSur samples yields a totally uninformative posterior on the kick. We
note that the NR waveform used here (with identifier
SXS:BBH:0137~\cite{SXSCatalog, Mroue:2013xna, Boyle:2019kee}) was not used to
train the surrogate models.

\prlsec{Kick measurement from existing GW events}
Next, we apply our method to GWTC-1 \cite{LIGOScientific:2018mvr} by reanalyzing the publicly available data
released by the LIGO-Virgo Collaborations~\cite{GW_open_science_center,
GWOSC:GWTC}. Figure~\ref{fig:GWTC_events} shows the posteriors we recover for
the kick magnitude for the GW150914~\cite{Abbott:2016blz} and
GW170729~\cite{LIGOScientific:2018mvr} events. These are compared with the prior for
the kick magnitude. Not much information about the kick can be gained for the
GWTC-1 events, as measured by the Kullback–Leibler (KL) divergence from the
prior to the posterior~\cite{Kullback_Leibler_divergence}. GW150914 and
GW170729 are those with the highest information gain showing respectively, a KL
divergence of 0.16 and 0.22 bits. This can be compared with
Ref.~\cite{LIGOScientific:2018mvr} where $\roughly0.13$ bits of information
gain in the precession parameter $\chi_p$~\cite{Schmidt:2014iyl} was considered
insufficient to claim evidence of precession. As an example of a good kick
measurement, the purple distribution in Fig.~\ref{fig:compare_models} has a KL
divergence of 1.74 bits with respect to the prior. While our kick measurement
for GW150914 is consistent with the 90\%-credible bound placed by
Ref.~\cite{Healy:2019} of $|\bv_f| \leq 492$ km/s, we find that this is driven
by the prior---meaning that that the measurement in Ref.~\cite{Healy:2019} was
largely uninformative.

Future detections will lead to much better constraints on the kick. In the
following sections, we explore the prospects for measuring kicks at the design
sensitivity of LIGO and Virgo.

\prlsec{Superkicks at design sensitivity}
We first consider a special binary BH configuration that is fine-tuned to
achieve a large kick velocity: both BHs have equal masses ($m_1=m_2=35\,
M_{\odot}$) and equal spin magnitudes ($|\bchi_1|=|\bchi_2|=0.5$); the spins
are entirely in the orbital plane and are antiparallel to each other at a
reference frequency $f_{\rm ref}=20$ Hz. The angle $\alpha$ between the
$x$-axis and the in-plane spins of the BHs  is allowed to vary. This
configuration is shown in the inset in the left panel of
Fig.~\ref{fig:superkick}. For concreteness, we choose luminosity distance
$d_L=440$ Mpc, inclination $\iota=0$ and orbital phase $\phi_{\rm ref}=0$.

\begin{figure}[tb]
\includegraphics[width=0.43\textwidth]{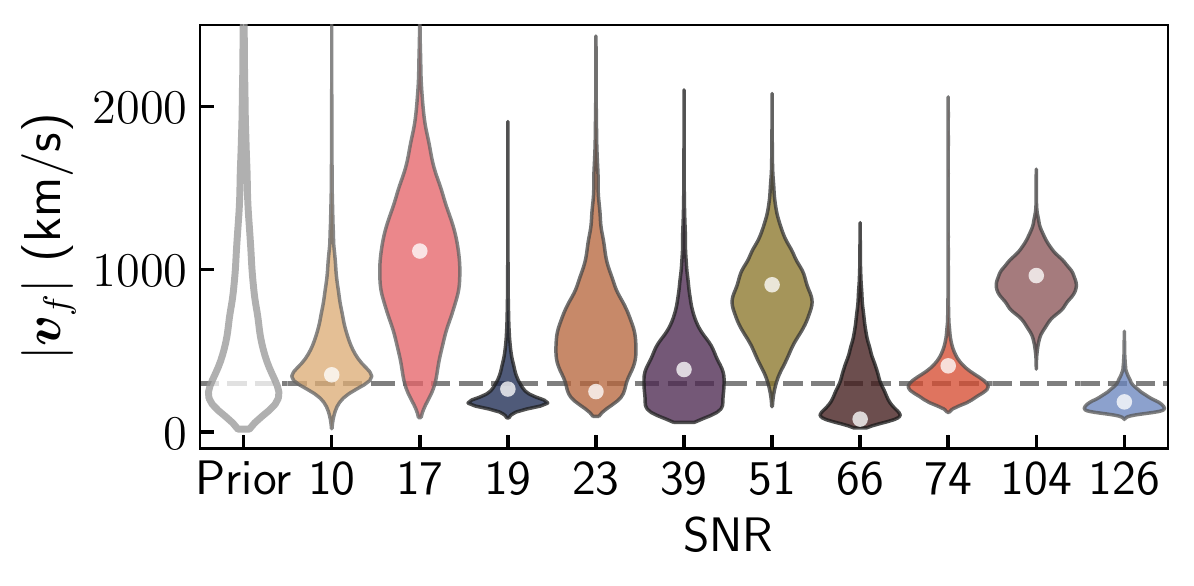}
\hfill\includegraphics[width=0.415\textwidth]{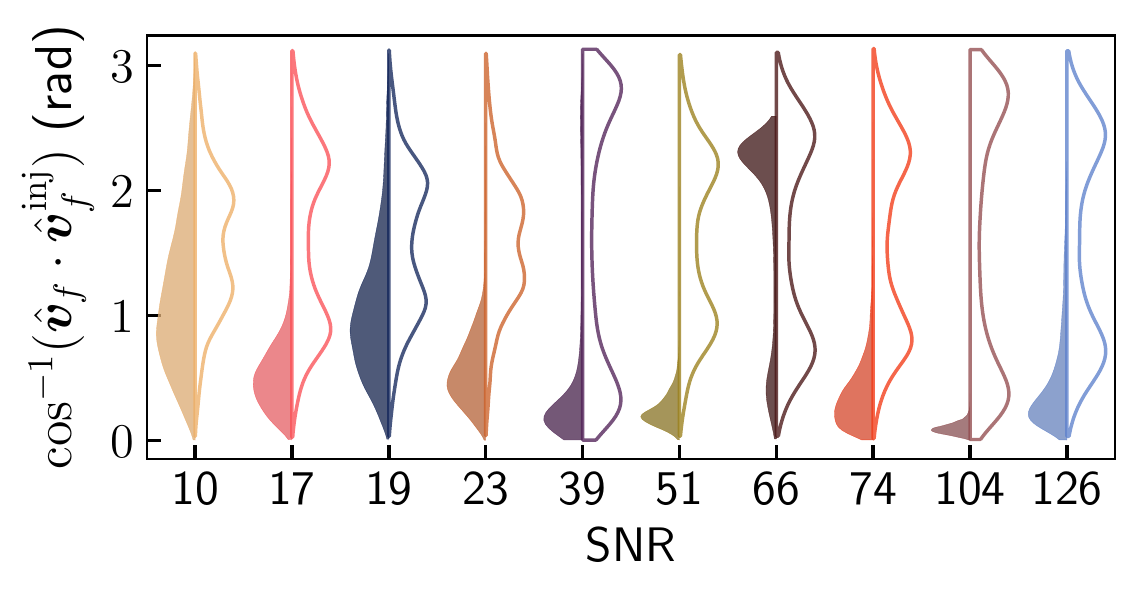}
\caption{Posteriors for the kick magnitude (top) and the kick-direction
    ``bias'' $\cos^{-1}(\hat{\bv}_f \cdot \hat{\bv}_f^{\rm inj})$ (bottom) for
    generic binary BH signals injected into design LIGO-Virgo noise. The
    probability distributions over the vertical axes are represented by full
    (half) violins in the top (bottom) panel, with thickness corresponding to
    probability density (normalized so all violins have equal width). In the
    top panel, the injected value is shown as a circular marker; in the
    bottom-panel, the injected value corresponds to zero bias. The recovered
    SNR is shown on the horizontal axes. The effective priors are represented
    by empty violins. The dashed gray line in the top-panel represents $|\bv_f|
    = 300$ km/s. For injected kick magnitudes below this line, \NRSurRemnant is
    known to be less accurate in predicting the kick direction (bottom).
}
\label{fig:random_cases}
\end{figure}

The kick magnitude has a sinusoidal dependence on $\alpha$~\cite{Lousto:2012su,
Brugmann:2007zj, Zlochower:2015wga, Gerosa:2018qay}, as shown in the left panel
of Fig.~\ref{fig:superkick} (see Ref.~\cite{Varma:2018rcg} for visualizations
of the sinusoidal dependence and superkicks.). We use \NRSurRemnant to find the
value of $\alpha$ that yields the maximum kick for the chosen spin magnitude.
We consider the $\alpha$ values that lead to the superkick ($|\bv_f|$=1814
km/s), half of the superkick ($|\bv_f|$=907 km/s), and a minimum kick magnitude
($|\bv_f|$=35 km/s)~\cite{kickpaperminkickfootnote}. The right-panel of
Fig.~\ref{fig:superkick} shows the kick magnitude posteriors obtained by
applying our method to \NRSur injections corresponding to those three
configurations. We are able to clearly distinguish the kick velocity
between these injections, which have otherwise nearly identical parameters.
This is in agreement with Ref.~\cite{Lousto:2019lyf}, where a mismatch
comparison was used to assess distinguishability between similar
configurations. The kick magnitude can be reliably recovered in all three
cases, demonstrating our ability to accurately measure superkicks at the design
sensitivity of LIGO and Virgo.

\begin{figure}[thb]
\includegraphics[width=0.47\textwidth]{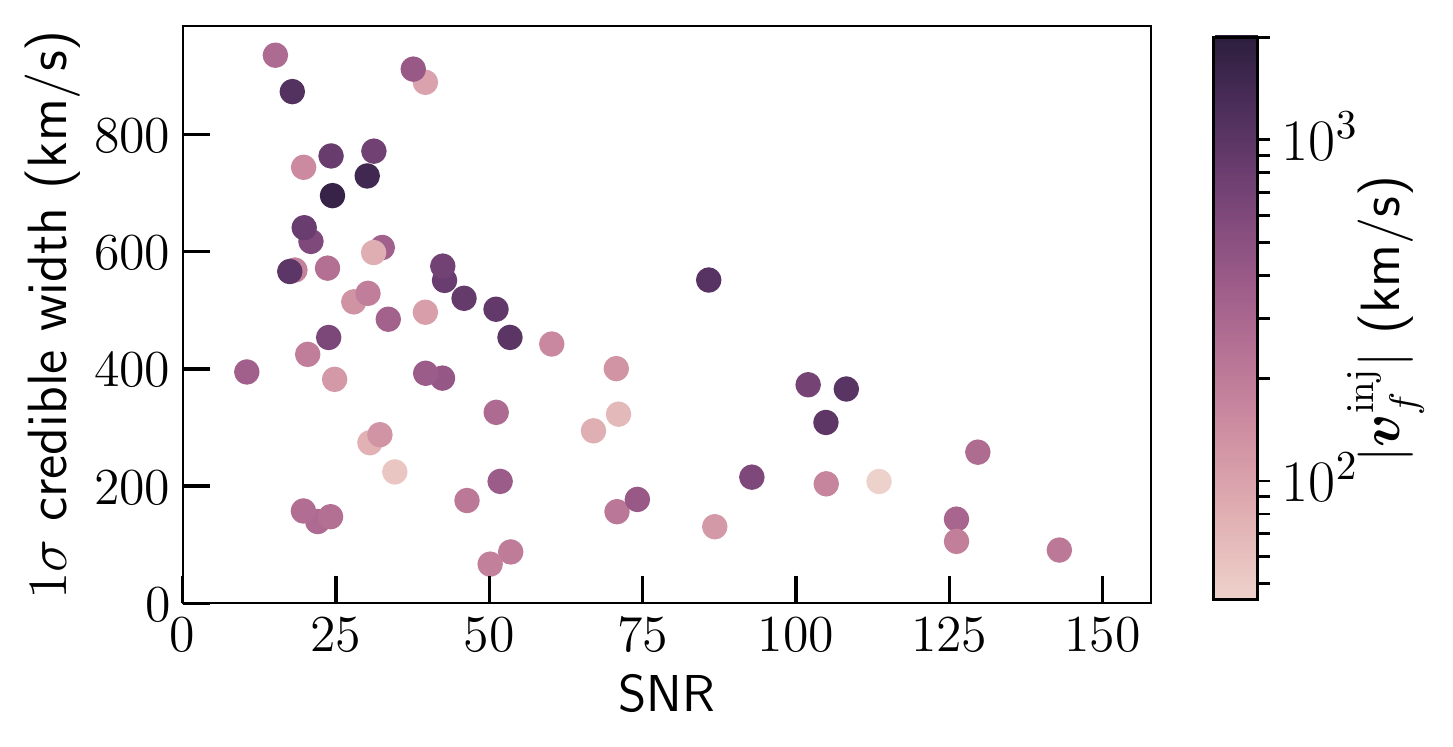}
\caption{Measurement uncertainty in the kick magnitude for randomly chosen
    binary BHs at the design sensitivity of LIGO and Virgo. The vertical axis
    shows the width of the shortest interval containing 68.27\%
    (${\sim}1\sigma$) of the posterior probability mass. Color indicates the
    injected kick magnitudes.
}
\label{fig:kick_vs_snr}
\end{figure}

\prlsec{Measuring kicks from generic systems}
The large kicks explored in the previous section required some fine-tuning of
the component parameters. For generic systems that are more likely to occur in
nature, typical kicks are much smaller~\cite{Berti:2012zp, Lousto:2012su}. We
now explore the measurability of the kick velocity of arbitrary systems by
injecting randomly chosen signals and studying the recovered kicks.  We perform
60 \NRSur injections uniformly sampled from mass ratios $q\in[1,3]$, spin
magnitudes $|\bchi_1|, |\bchi_2| \in [0,0.8]$, arbitrary spin directions, total
masses $M\in[70, 150]$, luminosity distances $d_L\in[400, 2000]$ Mpc,
inclination angles $\iota \in [0, \pi]$, and reference phases
$\phi_{\text{ref}} \in [0, 2\pi]$. These ranges are chosen to fall within the
training region of current surrogate models~\cite{Varma:2019csw}.

The recovered posteriors for the kick magnitude are shown in the top-panel of
Fig.~\ref{fig:random_cases} for a subset of 10 representative cases. Our method
reliably recovers the kick magnitude for these generic systems; biases away
from the true value are consistent with statistical error, as shown in the
Supplement \cite{kickpapersupplement}.

Figure~\ref{fig:kick_vs_snr} shows the measurement uncertainty in the recovered
kick magnitude for all 60 random cases. In general, a larger SNR leads to a
better measurement of the kick magnitude, but the specific choice of injected
parameters also plays a role, causing the spread in Fig.~\ref{fig:kick_vs_snr}.
In some cases a good measurement can be made at SNRs as low as 20.  This
suggests that kick velocities can be measured using our method even before LIGO
and Virgo achieve their design sensitivities.

Our method measures the full kick vector. To gauge how well we can recover the
kick direction, we consider the angle between the measured kick direction
$\hat{\bv}_f$ and the injected kick direction $\hat{\bv}_f^{\rm inj}$, namely
$\cos^{-1}(\hat{\bv}_f \cdot \hat{\bv}_f^{\rm inj})$. We refer to this angle as
the kick-direction ``bias''; for the true injection value, this angle is zero.
The bottom-panel of Fig.~\ref{fig:random_cases} shows the distribution of this
quantity as derived from the full kick-vector posteriors corresponding to the
same 10 cases as the top-panel. For all cases where the injected kick magnitude
is $\gtrsim 300$ km/s we recover the kick direction,
i.e.~$\cos^{-1}(\hat{\bv}_f \cdot \hat{\bv}_f^{\rm inj})\approx 0$.  For
smaller kick magnitudes, \NRSurRemnant is known to have larger intrinsic errors
in the kick direction~\cite{Varma:2019csw}, which results in correspondingly
higher kick-direction posterior biases. This comes from similar errors in the
underlying NR simulations on which the surrogate model is
trained~\cite{Varma:2019csw}, and should thus be fixed by more accurate
simulations. In spite of this, the kick magnitude is reliably recovered even
for cases with $|\bv_f| \lesssim 300$ km/s.

\prlsec{Applications}
Based on Fig.~\ref{fig:kick_vs_snr}, we generally expect an uncertainty of
$\lesssim 500$ km/s at SNR $\roughly 50$ in measuring the kick magnitude at the
68.27\% credible level (${\sim}1\sigma$). This can be used to place meaningful
constraints on the retention rate of the remnant for different types of
galaxies. For example, a kick measurement of the type shown in
Fig.~\ref{fig:compare_models} would lead us to conclude that the remnant of
such a binary would be ejected from most globular clusters, which typically
have escape velocities $\lesssim 50$ km/s~\cite{Antonini:2016gqe,
Merritt:2004xa}.

\begin{figure}[tb]
\includegraphics[width=0.41\textwidth]{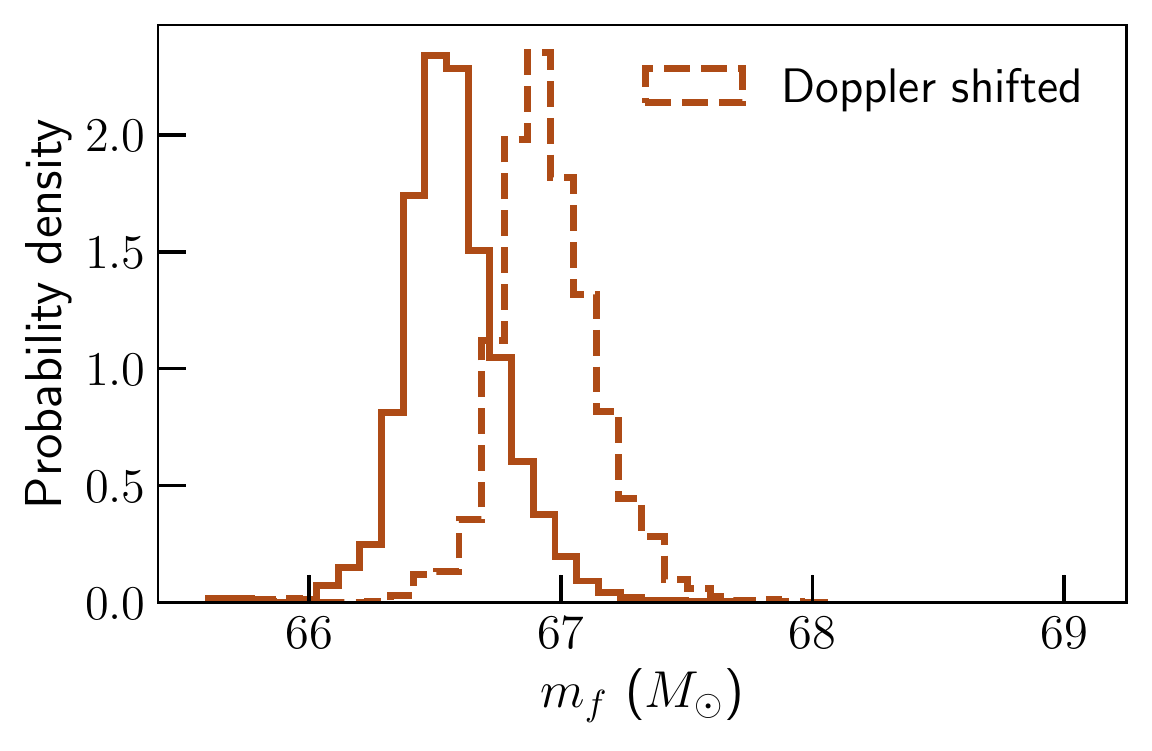}
\caption{The remnant mass and the Doppler-shifted remnant mass for the
    superkick configuration in Fig.~\ref{fig:superkick} ($\alpha=1.76$ rad).
    Not accounting for the expected difference in these distributions would
    result in a systematic bias in ringdown tests of general relativity, as
    detailed in the Supplement~\cite{kickpapersupplement}.
}
\label{fig:doppler_mass}
\end{figure}

In Fig.~\ref{fig:doppler_mass}, we use the projection of the full kick vector
along the line of sight to compute the kick's effect on the remnant BH
mass~\cite{Gerosa:2016vip} that would be inferred by an analysis of the
Doppler-shifted ringdown signal. As detectors become more sensitive, this
effect will need to be accounted for to avoid systematic biases in tests of
general relativity, especially for third-generation detectors and remnants with
large kick velocities along the line of sight. Our method will prevent these
issues, as we discuss in the Supplement~\cite{kickpapersupplement}.

\prlsec{Conclusion}
We present the first method to accurately extract both the kick magnitude and
direction of generically precessing binary BHs. This is made possible by recent
NR surrogate models for the gravitational waveform and properties of the merger
remnant (Fig.~\ref{fig:compare_models}).

We find that the SNR for existing GWTC-1 events is not sufficient to make a
confident measurement of the kick velocity (Fig.~\ref{fig:GWTC_events}).
However, our results indicate that the kick velocity will be reliably measured
once LIGO and Virgo reach their design sensitivities. This includes systems
with arbitrary parameters (Fig.~\ref{fig:random_cases}), as well as
configurations fine-tuned to produce superkicks with $|\bv_f| \sim 1000$ km/s
(Fig.~\ref{fig:superkick}). Measuring such kicks was previously estimated to be
only possible with third-generation GW detectors~\cite{Gerosa:2016vip}. On the
contrary, we find that accurate waveform and remnant surrogate models will soon
enable this with existing detectors (Fig.~\ref{fig:kick_vs_snr}).  This is in
agreement with Ref.~\cite{CalderonBustillo:2018zuq}, which made compatible
predictions for nonprecessing systems, for which $|\bv_f| \lesssim 300$ km/s.

Kick measurements obtained with our method can be used to place independent
constraints on the retention rate of the remnant BH in binary BH mergers, which
is directly related to the rate of second-generation mergers. In addition, we
show (Fig.~\ref{fig:doppler_mass} and Supplement~\cite{kickpapersupplement})
that kicks must be factored into ringdown tests of general relativity with
third-generation GW detectors to avoid systematic biases.

In this study, we focused on projected measurements by LIGO and Virgo at design
sensitivity. Since the kick velocity is very well recovered in some
moderate-SNR cases, we expect that our method may yield a successful kick
measurement before design sensitivity is achieved. This would mark the first
time a gravitational recoil is experimentally studied with GWs, providing a
brand new observable for astrophysics.

\prlsec{Acknowledgments}
We thank Juan Calderon Bustillo for a review and comments on the manuscript.
We thank Nathan Johnson-McDaniel, Ajith Parameswaran, Davide Gerosa, Matt
Giesler, Leo Stein, Saul Teukolsky, Gregorio Carullo, Aaron Zimmerman, and Bala
Iyer for useful discussions. V.V.\ is supported by the Sherman Fairchild
Foundation, and NSF grants PHY–170212 and PHY–1708213 at Caltech.
M.I.\ is supported by NASA through the NASA Hubble Fellowship
grant No.\ HST-HF2-51410.001-A awarded by the Space Telescope
Science Institute, which is operated by the Association of Universities
for Research in Astronomy, Inc., for NASA, under contract NAS5-26555.
This research made use of data, software and/or web tools obtained from the
Gravitational Wave Open Science Center~\cite{GW_open_science_center}, a service
of the LIGO Laboratory, the LIGO Scientific Collaboration and the Virgo
Collaboration.
Computations were performed on the Alice cluster at ICTS and the Wheeler
cluster at Caltech.
This paper carries LIGO document number \dcc{}.

\bibliography{References}

\begin{thebibliography}{99}%
\makeatletter
\providecommand \@ifxundefined [1]{%
 \@ifx{#1\undefined}
}%
\providecommand \@ifnum [1]{%
 \ifnum #1\expandafter \@firstoftwo
 \else \expandafter \@secondoftwo
 \fi
}%
\providecommand \@ifx [1]{%
 \ifx #1\expandafter \@firstoftwo
 \else \expandafter \@secondoftwo
 \fi
}%
\providecommand \natexlab [1]{#1}%
\providecommand \enquote  [1]{``#1''}%
\providecommand \bibnamefont  [1]{#1}%
\providecommand \bibfnamefont [1]{#1}%
\providecommand \citenamefont [1]{#1}%
\providecommand \href@noop [0]{\@secondoftwo}%
\providecommand \href [0]{\begingroup \@sanitize@url \@href}%
\providecommand \@href[1]{\@@startlink{#1}\@@href}%
\providecommand \@@href[1]{\endgroup#1\@@endlink}%
\providecommand \@sanitize@url [0]{\catcode `\\12\catcode `\$12\catcode
  `\&12\catcode `\#12\catcode `\^12\catcode `\_12\catcode `\%12\relax}%
\providecommand \@@startlink[1]{}%
\providecommand \@@endlink[0]{}%
\providecommand \url  [0]{\begingroup\@sanitize@url \@url }%
\providecommand \@url [1]{\endgroup\@href {#1}{\urlprefix }}%
\providecommand \urlprefix  [0]{URL }%
\providecommand \Eprint [0]{\href }%
\providecommand \doibase [0]{http://dx.doi.org/}%
\providecommand \selectlanguage [0]{\@gobble}%
\providecommand \bibinfo  [0]{\@secondoftwo}%
\providecommand \bibfield  [0]{\@secondoftwo}%
\providecommand \translation [1]{[#1]}%
\providecommand \BibitemOpen [0]{}%
\providecommand \bibitemStop [0]{}%
\providecommand \bibitemNoStop [0]{.\EOS\space}%
\providecommand \EOS [0]{\spacefactor3000\relax}%
\providecommand \BibitemShut  [1]{\csname bibitem#1\endcsname}%
\let\auto@bib@innerbib\@empty
\bibitem [{\citenamefont {Aasi}\ \emph {et~al.}(2015)\citenamefont {Aasi} \emph
  {et~al.}}]{TheLIGOScientific:2014jea}%
  \BibitemOpen
  \bibfield  {author} {\bibinfo {author} {\bibfnamefont {J.}~\bibnamefont
  {Aasi}} \emph {et~al.} (\bibinfo {collaboration} {LIGO Scientific}),\
  }\bibfield  {title} {\enquote {\bibinfo {title} {{Advanced LIGO}},}\ }\href
  {\doibase 10.1088/0264-9381/32/7/074001} {\bibfield  {journal} {\bibinfo
  {journal} {Class. Quant. Grav.}\ }\textbf {\bibinfo {volume} {32}},\ \bibinfo
  {pages} {074001} (\bibinfo {year} {2015})},\ \Eprint
  {http://arxiv.org/abs/1411.4547} {arXiv:1411.4547 [gr-qc]} \BibitemShut
  {NoStop}%
\bibitem [{\citenamefont {Acernese}\ \emph {et~al.}(2015)\citenamefont
  {Acernese} \emph {et~al.}}]{TheVirgo:2014hva}%
  \BibitemOpen
  \bibfield  {author} {\bibinfo {author} {\bibfnamefont {F.}~\bibnamefont
  {Acernese}} \emph {et~al.} (\bibinfo {collaboration} {Virgo}),\ }\bibfield
  {title} {\enquote {\bibinfo {title} {{Advanced Virgo: a second-generation
  interferometric gravitational wave detector}},}\ }\href {\doibase
  10.1088/0264-9381/32/2/024001} {\bibfield  {journal} {\bibinfo  {journal}
  {Class. Quant. Grav.}\ }\textbf {\bibinfo {volume} {32}},\ \bibinfo {pages}
  {024001} (\bibinfo {year} {2015})},\ \Eprint {http://arxiv.org/abs/1408.3978}
  {arXiv:1408.3978 [gr-qc]} \BibitemShut {NoStop}%
\bibitem [{\citenamefont {Bonnor}\ \emph {et~al.}(1961)\citenamefont {Bonnor},
  \citenamefont {Rotenberg},\ and\ \citenamefont {Louis}}]{Bonnor:1961linmom}%
  \BibitemOpen
  \bibfield  {author} {\bibinfo {author} {\bibfnamefont {W.~B.}\ \bibnamefont
  {Bonnor}}, \bibinfo {author} {\bibfnamefont {M.~A.}\ \bibnamefont
  {Rotenberg}}, \ and\ \bibinfo {author} {\bibfnamefont {Rosenhead}\
  \bibnamefont {Louis}},\ }\bibfield  {title} {\enquote {\bibinfo {title}
  {{Transport of momentum by gravitational waves: the linear approximation}},}\
  }\href {\doibase 10.1098/rspa.1961.0226} {\bibfield  {journal} {\bibinfo
  {journal} {Proceedings of the Royal Society of London Series A.}\ }\textbf
  {\bibinfo {volume} {265}} (\bibinfo {year} {1961}),\
  10.1098/rspa.1961.0226}\BibitemShut {NoStop}%
\bibitem [{\citenamefont {Peres}(1962)}]{PhysRev.128.2471}%
  \BibitemOpen
  \bibfield  {author} {\bibinfo {author} {\bibfnamefont {Asher}\ \bibnamefont
  {Peres}},\ }\bibfield  {title} {\enquote {\bibinfo {title} {Classical
  radiation recoil},}\ }\href {\doibase 10.1103/PhysRev.128.2471} {\bibfield
  {journal} {\bibinfo  {journal} {Phys. Rev.}\ }\textbf {\bibinfo {volume}
  {128}},\ \bibinfo {pages} {2471--2475} (\bibinfo {year} {1962})}\BibitemShut
  {NoStop}%
\bibitem [{\citenamefont {{Bekenstein}}(1973)}]{Bekenstein:1973ApJ}%
  \BibitemOpen
  \bibfield  {author} {\bibinfo {author} {\bibfnamefont {Jacob~D.}\
  \bibnamefont {{Bekenstein}}},\ }\bibfield  {title} {\enquote {\bibinfo
  {title} {{Gravitational-Radiation Recoil and Runaway Black Holes}},}\ }\href
  {\doibase 10.1086/152255} {\bibfield  {journal} {\bibinfo  {journal} {The
  Astrophysical Journal}\ }\textbf {\bibinfo {volume} {183}},\ \bibinfo {pages}
  {657--664} (\bibinfo {year} {1973})}\BibitemShut {NoStop}%
\bibitem [{\citenamefont {Fitchett}(1983)}]{Fitchett:1983MNRAS}%
  \BibitemOpen
  \bibfield  {author} {\bibinfo {author} {\bibfnamefont {M.~J.}\ \bibnamefont
  {Fitchett}},\ }\bibfield  {title} {\enquote {\bibinfo {title} {{The influence
  of gravitational wave momentum losses on the centre of mass motion of a
  Newtonian binary system}},}\ }\href {\doibase 10.1093/mnras/203.4.1049}
  {\bibfield  {journal} {\bibinfo  {journal} {Monthly Notices of the Royal
  Astronomical Society}\ }\textbf {\bibinfo {volume} {203}},\ \bibinfo {pages}
  {1049--1062} (\bibinfo {year} {1983})},\ \Eprint
  {http://arxiv.org/abs/http://oup.prod.sis.lan/mnras/article-pdf/203/4/1049/18223796/mnras203-1049.pdf}
  {http://oup.prod.sis.lan/mnras/article-pdf/203/4/1049/18223796/mnras203-1049.pdf}
  \BibitemShut {NoStop}%
\bibitem [{\citenamefont {Gonzalez}\ \emph
  {et~al.}(2007{\natexlab{a}})\citenamefont {Gonzalez}, \citenamefont
  {Sperhake}, \citenamefont {Bruegmann}, \citenamefont {Hannam},\ and\
  \citenamefont {Husa}}]{Gonzalez:2006md}%
  \BibitemOpen
  \bibfield  {author} {\bibinfo {author} {\bibfnamefont {Jose~A.}\ \bibnamefont
  {Gonzalez}}, \bibinfo {author} {\bibfnamefont {Ulrich}\ \bibnamefont
  {Sperhake}}, \bibinfo {author} {\bibfnamefont {Bernd}\ \bibnamefont
  {Bruegmann}}, \bibinfo {author} {\bibfnamefont {Mark}\ \bibnamefont
  {Hannam}}, \ and\ \bibinfo {author} {\bibfnamefont {Sascha}\ \bibnamefont
  {Husa}},\ }\bibfield  {title} {\enquote {\bibinfo {title} {{Total recoil: The
  Maximum kick from nonspinning black-hole binary inspiral}},}\ }\href
  {\doibase 10.1103/PhysRevLett.98.091101} {\bibfield  {journal} {\bibinfo
  {journal} {Phys. Rev. Lett.}\ }\textbf {\bibinfo {volume} {98}},\ \bibinfo
  {pages} {091101} (\bibinfo {year} {2007}{\natexlab{a}})},\ \Eprint
  {http://arxiv.org/abs/gr-qc/0610154} {arXiv:gr-qc/0610154 [gr-qc]}
  \BibitemShut {NoStop}%
\bibitem [{\citenamefont {Lousto}\ and\ \citenamefont
  {Zlochower}(2008)}]{Lousto:2007db}%
  \BibitemOpen
  \bibfield  {author} {\bibinfo {author} {\bibfnamefont {Carlos~O.}\
  \bibnamefont {Lousto}}\ and\ \bibinfo {author} {\bibfnamefont {Yosef}\
  \bibnamefont {Zlochower}},\ }\bibfield  {title} {\enquote {\bibinfo {title}
  {{Further insight into gravitational recoil}},}\ }\href {\doibase
  10.1103/PhysRevD.77.044028} {\bibfield  {journal} {\bibinfo  {journal} {Phys.
  Rev.}\ }\textbf {\bibinfo {volume} {D77}},\ \bibinfo {pages} {044028}
  (\bibinfo {year} {2008})},\ \Eprint {http://arxiv.org/abs/0708.4048}
  {arXiv:0708.4048 [gr-qc]} \BibitemShut {NoStop}%
\bibitem [{\citenamefont {Lousto}\ \emph {et~al.}(2012)\citenamefont {Lousto},
  \citenamefont {Zlochower}, \citenamefont {Dotti},\ and\ \citenamefont
  {Volonteri}}]{Lousto:2012su}%
  \BibitemOpen
  \bibfield  {author} {\bibinfo {author} {\bibfnamefont {Carlos~O.}\
  \bibnamefont {Lousto}}, \bibinfo {author} {\bibfnamefont {Yosef}\
  \bibnamefont {Zlochower}}, \bibinfo {author} {\bibfnamefont {Massimo}\
  \bibnamefont {Dotti}}, \ and\ \bibinfo {author} {\bibfnamefont {Marta}\
  \bibnamefont {Volonteri}},\ }\bibfield  {title} {\enquote {\bibinfo {title}
  {{Gravitational Recoil From Accretion-Aligned Black-Hole Binaries}},}\ }\href
  {\doibase 10.1103/PhysRevD.85.084015} {\bibfield  {journal} {\bibinfo
  {journal} {Phys. Rev.}\ }\textbf {\bibinfo {volume} {D85}},\ \bibinfo {pages}
  {084015} (\bibinfo {year} {2012})},\ \Eprint {http://arxiv.org/abs/1201.1923}
  {arXiv:1201.1923 [gr-qc]} \BibitemShut {NoStop}%
\bibitem [{\citenamefont {Lousto}\ and\ \citenamefont
  {Zlochower}(2013)}]{Lousto:2012gt}%
  \BibitemOpen
  \bibfield  {author} {\bibinfo {author} {\bibfnamefont {Carlos~O.}\
  \bibnamefont {Lousto}}\ and\ \bibinfo {author} {\bibfnamefont {Yosef}\
  \bibnamefont {Zlochower}},\ }\bibfield  {title} {\enquote {\bibinfo {title}
  {{Nonlinear Gravitational Recoil from the Mergers of Precessing Black-Hole
  Binaries}},}\ }\href {\doibase 10.1103/PhysRevD.87.084027} {\bibfield
  {journal} {\bibinfo  {journal} {Phys. Rev.}\ }\textbf {\bibinfo {volume}
  {D87}},\ \bibinfo {pages} {084027} (\bibinfo {year} {2013})},\ \Eprint
  {http://arxiv.org/abs/1211.7099} {arXiv:1211.7099 [gr-qc]} \BibitemShut
  {NoStop}%
\bibitem [{\citenamefont {Blanchet}\ \emph {et~al.}(2005)\citenamefont
  {Blanchet}, \citenamefont {Qusailah},\ and\ \citenamefont
  {Will}}]{Blanchet:2005rj}%
  \BibitemOpen
  \bibfield  {author} {\bibinfo {author} {\bibfnamefont {Luc}\ \bibnamefont
  {Blanchet}}, \bibinfo {author} {\bibfnamefont {Moh'd S.~S.}\ \bibnamefont
  {Qusailah}}, \ and\ \bibinfo {author} {\bibfnamefont {Clifford~M.}\
  \bibnamefont {Will}},\ }\bibfield  {title} {\enquote {\bibinfo {title}
  {{Gravitational recoil of inspiralling black-hole binaries to second
  post-Newtonian order}},}\ }\href {\doibase 10.1086/497332} {\bibfield
  {journal} {\bibinfo  {journal} {Astrophys. J.}\ }\textbf {\bibinfo {volume}
  {635}},\ \bibinfo {pages} {508} (\bibinfo {year} {2005})},\ \Eprint
  {http://arxiv.org/abs/astro-ph/0507692} {arXiv:astro-ph/0507692 [astro-ph]}
  \BibitemShut {NoStop}%
\bibitem [{\citenamefont {Damour}\ and\ \citenamefont
  {Gopakumar}(2006)}]{Damour:2006tr}%
  \BibitemOpen
  \bibfield  {author} {\bibinfo {author} {\bibfnamefont {Thibault}\
  \bibnamefont {Damour}}\ and\ \bibinfo {author} {\bibfnamefont {Achamveedu}\
  \bibnamefont {Gopakumar}},\ }\bibfield  {title} {\enquote {\bibinfo {title}
  {{Gravitational recoil during binary black hole coalescence using the
  effective one body approach}},}\ }\href {\doibase 10.1103/PhysRevD.73.124006}
  {\bibfield  {journal} {\bibinfo  {journal} {Phys. Rev.}\ }\textbf {\bibinfo
  {volume} {D73}},\ \bibinfo {pages} {124006} (\bibinfo {year} {2006})},\
  \Eprint {http://arxiv.org/abs/gr-qc/0602117} {arXiv:gr-qc/0602117 [gr-qc]}
  \BibitemShut {NoStop}%
\bibitem [{\citenamefont {Le~Tiec}\ \emph {et~al.}(2010)\citenamefont
  {Le~Tiec}, \citenamefont {Blanchet},\ and\ \citenamefont
  {Will}}]{LeTiec:2009yg}%
  \BibitemOpen
  \bibfield  {author} {\bibinfo {author} {\bibfnamefont {Alexandre}\
  \bibnamefont {Le~Tiec}}, \bibinfo {author} {\bibfnamefont {Luc}\ \bibnamefont
  {Blanchet}}, \ and\ \bibinfo {author} {\bibfnamefont {Clifford~M.}\
  \bibnamefont {Will}},\ }\bibfield  {title} {\enquote {\bibinfo {title}
  {{Gravitational-Wave Recoil from the Ringdown Phase of Coalescing Black Hole
  Binaries}},}\ }\href {\doibase 10.1088/0264-9381/27/1/012001} {\bibfield
  {journal} {\bibinfo  {journal} {Class. Quant. Grav.}\ }\textbf {\bibinfo
  {volume} {27}},\ \bibinfo {pages} {012001} (\bibinfo {year} {2010})},\
  \Eprint {http://arxiv.org/abs/0910.4594} {arXiv:0910.4594 [gr-qc]}
  \BibitemShut {NoStop}%
\bibitem [{\citenamefont {{Israel}}(1968)}]{1968_Israel}%
  \BibitemOpen
  \bibfield  {author} {\bibinfo {author} {\bibfnamefont {W.}~\bibnamefont
  {{Israel}}},\ }\bibfield  {title} {\enquote {\bibinfo {title} {{Event
  horizons in static electrovac space-times}},}\ }\href {\doibase
  10.1007/BF01645859} {\bibfield  {journal} {\bibinfo  {journal}
  {Communications in Mathematical Physics}\ }\textbf {\bibinfo {volume} {8}},\
  \bibinfo {pages} {245--260} (\bibinfo {year} {1968})}\BibitemShut {NoStop}%
\bibitem [{\citenamefont {Carter}(1971)}]{PhysRevLett.26.331}%
  \BibitemOpen
  \bibfield  {author} {\bibinfo {author} {\bibfnamefont {B.}~\bibnamefont
  {Carter}},\ }\bibfield  {title} {\enquote {\bibinfo {title} {Axisymmetric
  black hole has only two degrees of freedom},}\ }\href {\doibase
  10.1103/PhysRevLett.26.331} {\bibfield  {journal} {\bibinfo  {journal} {Phys.
  Rev. Lett.}\ }\textbf {\bibinfo {volume} {26}},\ \bibinfo {pages} {331--333}
  (\bibinfo {year} {1971})}\BibitemShut {NoStop}%
\bibitem [{\citenamefont {Abbott}\ \emph
  {et~al.}(2016{\natexlab{a}})\citenamefont {Abbott} \emph
  {et~al.}}]{TheLIGOScientific:2016src}%
  \BibitemOpen
  \bibfield  {author} {\bibinfo {author} {\bibfnamefont {B.~P.}\ \bibnamefont
  {Abbott}} \emph {et~al.} (\bibinfo {collaboration} {LIGO Scientific,
  Virgo}),\ }\bibfield  {title} {\enquote {\bibinfo {title} {{Tests of general
  relativity with GW150914}},}\ }\href {\doibase
  10.1103/PhysRevLett.116.221101} {\bibfield  {journal} {\bibinfo  {journal}
  {Phys. Rev. Lett.}\ }\textbf {\bibinfo {volume} {116}},\ \bibinfo {pages}
  {221101} (\bibinfo {year} {2016}{\natexlab{a}})},\ \bibinfo {note} {[Erratum:
  Phys. Rev. Lett.121,no.12,129902(2018)]},\ \Eprint
  {http://arxiv.org/abs/1602.03841} {arXiv:1602.03841 [gr-qc]} \BibitemShut
  {NoStop}%
\bibitem [{\citenamefont {Abbott}\ \emph
  {et~al.}(2019{\natexlab{a}})\citenamefont {Abbott} \emph
  {et~al.}}]{LIGOScientific:2019fpa}%
  \BibitemOpen
  \bibfield  {author} {\bibinfo {author} {\bibfnamefont {B.~P.}\ \bibnamefont
  {Abbott}} \emph {et~al.} (\bibinfo {collaboration} {LIGO Scientific,
  Virgo}),\ }\bibfield  {title} {\enquote {\bibinfo {title} {{Tests of General
  Relativity with the Binary Black Hole Signals from the LIGO-Virgo Catalog
  GWTC-1}},}\ }\href {\doibase 10.1103/PhysRevD.100.104036} {\bibfield
  {journal} {\bibinfo  {journal} {Phys. Rev.}\ }\textbf {\bibinfo {volume}
  {D100}},\ \bibinfo {pages} {104036} (\bibinfo {year} {2019}{\natexlab{a}})},\
  \Eprint {http://arxiv.org/abs/1903.04467} {arXiv:1903.04467 [gr-qc]}
  \BibitemShut {NoStop}%
\bibitem [{\citenamefont {Ghosh}\ \emph {et~al.}(2018)\citenamefont {Ghosh},
  \citenamefont {Johnson-Mcdaniel}, \citenamefont {Ghosh}, \citenamefont
  {Mishra}, \citenamefont {Ajith}, \citenamefont {Del~Pozzo}, \citenamefont
  {Berry}, \citenamefont {Nielsen},\ and\ \citenamefont
  {London}}]{Ghosh:2017gfp}%
  \BibitemOpen
  \bibfield  {author} {\bibinfo {author} {\bibfnamefont {Abhirup}\ \bibnamefont
  {Ghosh}}, \bibinfo {author} {\bibfnamefont {Nathan~K.}\ \bibnamefont
  {Johnson-Mcdaniel}}, \bibinfo {author} {\bibfnamefont {Archisman}\
  \bibnamefont {Ghosh}}, \bibinfo {author} {\bibfnamefont {Chandra~Kant}\
  \bibnamefont {Mishra}}, \bibinfo {author} {\bibfnamefont {Parameswaran}\
  \bibnamefont {Ajith}}, \bibinfo {author} {\bibfnamefont {Walter}\
  \bibnamefont {Del~Pozzo}}, \bibinfo {author} {\bibfnamefont {Christopher
  P.~L.}\ \bibnamefont {Berry}}, \bibinfo {author} {\bibfnamefont {Alex~B.}\
  \bibnamefont {Nielsen}}, \ and\ \bibinfo {author} {\bibfnamefont {Lionel}\
  \bibnamefont {London}},\ }\bibfield  {title} {\enquote {\bibinfo {title}
  {{Testing general relativity using gravitational wave signals from the
  inspiral, merger and ringdown of binary black holes}},}\ }\href {\doibase
  10.1088/1361-6382/aa972e} {\bibfield  {journal} {\bibinfo  {journal} {Class.
  Quant. Grav.}\ }\textbf {\bibinfo {volume} {35}},\ \bibinfo {pages} {014002}
  (\bibinfo {year} {2018})},\ \Eprint {http://arxiv.org/abs/1704.06784}
  {arXiv:1704.06784 [gr-qc]} \BibitemShut {NoStop}%
\bibitem [{\citenamefont {Brito}\ \emph {et~al.}(2018)\citenamefont {Brito},
  \citenamefont {Buonanno},\ and\ \citenamefont {Raymond}}]{Brito:2018rfr}%
  \BibitemOpen
  \bibfield  {author} {\bibinfo {author} {\bibfnamefont {Richard}\ \bibnamefont
  {Brito}}, \bibinfo {author} {\bibfnamefont {Alessandra}\ \bibnamefont
  {Buonanno}}, \ and\ \bibinfo {author} {\bibfnamefont {Vivien}\ \bibnamefont
  {Raymond}},\ }\bibfield  {title} {\enquote {\bibinfo {title} {{Black-hole
  Spectroscopy by Making Full Use of Gravitational-Wave Modeling}},}\ }\href
  {\doibase 10.1103/PhysRevD.98.084038} {\bibfield  {journal} {\bibinfo
  {journal} {Phys. Rev.}\ }\textbf {\bibinfo {volume} {D98}},\ \bibinfo {pages}
  {084038} (\bibinfo {year} {2018})},\ \Eprint
  {http://arxiv.org/abs/1805.00293} {arXiv:1805.00293 [gr-qc]} \BibitemShut
  {NoStop}%
\bibitem [{\citenamefont {Carullo}\ \emph {et~al.}(2018)\citenamefont {Carullo}
  \emph {et~al.}}]{Carullo:2018sfu}%
  \BibitemOpen
  \bibfield  {author} {\bibinfo {author} {\bibfnamefont {Gregorio}\
  \bibnamefont {Carullo}} \emph {et~al.},\ }\bibfield  {title} {\enquote
  {\bibinfo {title} {{Empirical tests of the black hole no-hair conjecture
  using gravitational-wave observations}},}\ }\href {\doibase
  10.1103/PhysRevD.98.104020} {\bibfield  {journal} {\bibinfo  {journal} {Phys.
  Rev.}\ }\textbf {\bibinfo {volume} {D98}},\ \bibinfo {pages} {104020}
  (\bibinfo {year} {2018})},\ \Eprint {http://arxiv.org/abs/1805.04760}
  {arXiv:1805.04760 [gr-qc]} \BibitemShut {NoStop}%
\bibitem [{\citenamefont {Carullo}\ \emph {et~al.}(2019)\citenamefont
  {Carullo}, \citenamefont {Del~Pozzo},\ and\ \citenamefont
  {Veitch}}]{Carullo:2019flw}%
  \BibitemOpen
  \bibfield  {author} {\bibinfo {author} {\bibfnamefont {Gregorio}\
  \bibnamefont {Carullo}}, \bibinfo {author} {\bibfnamefont {Walter}\
  \bibnamefont {Del~Pozzo}}, \ and\ \bibinfo {author} {\bibfnamefont {John}\
  \bibnamefont {Veitch}},\ }\bibfield  {title} {\enquote {\bibinfo {title}
  {{Observational Black Hole Spectroscopy: A time-domain multimode analysis of
  GW150914}},}\ }\href {\doibase 10.1103/PhysRevD.99.123029,
  10.1103/PhysRevD.100.089903} {\bibfield  {journal} {\bibinfo  {journal}
  {Phys. Rev.}\ }\textbf {\bibinfo {volume} {D99}},\ \bibinfo {pages} {123029}
  (\bibinfo {year} {2019})},\ \bibinfo {note} {[Erratum: Phys.
  Rev.D100,no.8,089903(2019)]},\ \Eprint {http://arxiv.org/abs/1902.07527}
  {arXiv:1902.07527 [gr-qc]} \BibitemShut {NoStop}%
\bibitem [{\citenamefont {Isi}\ \emph {et~al.}(2019)\citenamefont {Isi},
  \citenamefont {Giesler}, \citenamefont {Farr}, \citenamefont {Scheel},\ and\
  \citenamefont {Teukolsky}}]{Isi:2019aib}%
  \BibitemOpen
  \bibfield  {author} {\bibinfo {author} {\bibfnamefont {Maximiliano}\
  \bibnamefont {Isi}}, \bibinfo {author} {\bibfnamefont {Matthew}\ \bibnamefont
  {Giesler}}, \bibinfo {author} {\bibfnamefont {Will~M.}\ \bibnamefont {Farr}},
  \bibinfo {author} {\bibfnamefont {Mark~A.}\ \bibnamefont {Scheel}}, \ and\
  \bibinfo {author} {\bibfnamefont {Saul~A.}\ \bibnamefont {Teukolsky}},\
  }\bibfield  {title} {\enquote {\bibinfo {title} {{Testing the no-hair theorem
  with GW150914}},}\ }\href {\doibase 10.1103/PhysRevLett.123.111102}
  {\bibfield  {journal} {\bibinfo  {journal} {Phys. Rev. Lett.}\ }\textbf
  {\bibinfo {volume} {123}},\ \bibinfo {pages} {111102} (\bibinfo {year}
  {2019})},\ \Eprint {http://arxiv.org/abs/1905.00869} {arXiv:1905.00869
  [gr-qc]} \BibitemShut {NoStop}%
\bibitem [{\citenamefont {Giesler}\ \emph {et~al.}(2019)\citenamefont
  {Giesler}, \citenamefont {Isi}, \citenamefont {Scheel},\ and\ \citenamefont
  {Teukolsky}}]{Giesler:2019uxc}%
  \BibitemOpen
  \bibfield  {author} {\bibinfo {author} {\bibfnamefont {Matthew}\ \bibnamefont
  {Giesler}}, \bibinfo {author} {\bibfnamefont {Maximiliano}\ \bibnamefont
  {Isi}}, \bibinfo {author} {\bibfnamefont {Mark}\ \bibnamefont {Scheel}}, \
  and\ \bibinfo {author} {\bibfnamefont {Saul}\ \bibnamefont {Teukolsky}},\
  }\bibfield  {title} {\enquote {\bibinfo {title} {{Black hole ringdown: the
  importance of overtones}},}\ }\href {\doibase 10.1103/PhysRevX.9.041060}
  {\bibfield  {journal} {\bibinfo  {journal} {Phys. Rev.}\ }\textbf {\bibinfo
  {volume} {X9}},\ \bibinfo {pages} {041060} (\bibinfo {year} {2019})},\
  \Eprint {http://arxiv.org/abs/1903.08284} {arXiv:1903.08284 [gr-qc]}
  \BibitemShut {NoStop}%
\bibitem [{\citenamefont {Apostolatos}\ \emph {et~al.}(1994)\citenamefont
  {Apostolatos}, \citenamefont {Cutler}, \citenamefont {Sussman},\ and\
  \citenamefont {Thorne}}]{Apostolatos:1994pre}%
  \BibitemOpen
  \bibfield  {author} {\bibinfo {author} {\bibfnamefont {Theocharis~A.}\
  \bibnamefont {Apostolatos}}, \bibinfo {author} {\bibfnamefont {Curt}\
  \bibnamefont {Cutler}}, \bibinfo {author} {\bibfnamefont {Gerald~J.}\
  \bibnamefont {Sussman}}, \ and\ \bibinfo {author} {\bibfnamefont {Kip~S.}\
  \bibnamefont {Thorne}},\ }\bibfield  {title} {\enquote {\bibinfo {title}
  {Spin-induced orbital precession and its modulation of the gravitational
  waveforms from merging binaries},}\ }\href {\doibase
  10.1103/PhysRevD.49.6274} {\bibfield  {journal} {\bibinfo  {journal} {Phys.
  Rev. D}\ }\textbf {\bibinfo {volume} {49}},\ \bibinfo {pages} {6274--6297}
  (\bibinfo {year} {1994})}\BibitemShut {NoStop}%
\bibitem [{\citenamefont {Campanelli}\ \emph
  {et~al.}(2007{\natexlab{a}})\citenamefont {Campanelli}, \citenamefont
  {Lousto}, \citenamefont {Zlochower},\ and\ \citenamefont
  {Merritt}}]{Campanelli:2007cga}%
  \BibitemOpen
  \bibfield  {author} {\bibinfo {author} {\bibfnamefont {Manuela}\ \bibnamefont
  {Campanelli}}, \bibinfo {author} {\bibfnamefont {Carlos~O.}\ \bibnamefont
  {Lousto}}, \bibinfo {author} {\bibfnamefont {Yosef}\ \bibnamefont
  {Zlochower}}, \ and\ \bibinfo {author} {\bibfnamefont {David}\ \bibnamefont
  {Merritt}},\ }\bibfield  {title} {\enquote {\bibinfo {title} {{Maximum
  gravitational recoil}},}\ }\href {\doibase 10.1103/PhysRevLett.98.231102}
  {\bibfield  {journal} {\bibinfo  {journal} {Phys. Rev. Lett.}\ }\textbf
  {\bibinfo {volume} {98}},\ \bibinfo {pages} {231102} (\bibinfo {year}
  {2007}{\natexlab{a}})},\ \Eprint {http://arxiv.org/abs/gr-qc/0702133}
  {arXiv:gr-qc/0702133 [GR-QC]} \BibitemShut {NoStop}%
\bibitem [{\citenamefont {Gonzalez}\ \emph
  {et~al.}(2007{\natexlab{b}})\citenamefont {Gonzalez}, \citenamefont {Hannam},
  \citenamefont {Sperhake}, \citenamefont {Bruegmann},\ and\ \citenamefont
  {Husa}}]{Gonzalez:2007hi}%
  \BibitemOpen
  \bibfield  {author} {\bibinfo {author} {\bibfnamefont {J.~A.}\ \bibnamefont
  {Gonzalez}}, \bibinfo {author} {\bibfnamefont {M.~D.}\ \bibnamefont
  {Hannam}}, \bibinfo {author} {\bibfnamefont {U.}~\bibnamefont {Sperhake}},
  \bibinfo {author} {\bibfnamefont {Bernd}\ \bibnamefont {Bruegmann}}, \ and\
  \bibinfo {author} {\bibfnamefont {S.}~\bibnamefont {Husa}},\ }\bibfield
  {title} {\enquote {\bibinfo {title} {{Supermassive recoil velocities for
  binary black-hole mergers with antialigned spins}},}\ }\href {\doibase
  10.1103/PhysRevLett.98.231101} {\bibfield  {journal} {\bibinfo  {journal}
  {Phys. Rev. Lett.}\ }\textbf {\bibinfo {volume} {98}},\ \bibinfo {pages}
  {231101} (\bibinfo {year} {2007}{\natexlab{b}})},\ \Eprint
  {http://arxiv.org/abs/gr-qc/0702052} {arXiv:gr-qc/0702052 [GR-QC]}
  \BibitemShut {NoStop}%
\bibitem [{\citenamefont {Tichy}\ and\ \citenamefont
  {Marronetti}(2007)}]{Tichy:2007hk}%
  \BibitemOpen
  \bibfield  {author} {\bibinfo {author} {\bibfnamefont {Wolfgang}\
  \bibnamefont {Tichy}}\ and\ \bibinfo {author} {\bibfnamefont {Pedro}\
  \bibnamefont {Marronetti}},\ }\bibfield  {title} {\enquote {\bibinfo {title}
  {{Binary black hole mergers: Large kicks for generic spin orientations}},}\
  }\href {\doibase 10.1103/PhysRevD.76.061502} {\bibfield  {journal} {\bibinfo
  {journal} {Phys. Rev.}\ }\textbf {\bibinfo {volume} {D76}},\ \bibinfo {pages}
  {061502} (\bibinfo {year} {2007})},\ \Eprint
  {http://arxiv.org/abs/gr-qc/0703075} {arXiv:gr-qc/0703075 [gr-qc]}
  \BibitemShut {NoStop}%
\bibitem [{\citenamefont {Lousto}\ and\ \citenamefont
  {Zlochower}(2011)}]{Lousto:2011kp}%
  \BibitemOpen
  \bibfield  {author} {\bibinfo {author} {\bibfnamefont {Carlos~O.}\
  \bibnamefont {Lousto}}\ and\ \bibinfo {author} {\bibfnamefont {Yosef}\
  \bibnamefont {Zlochower}},\ }\bibfield  {title} {\enquote {\bibinfo {title}
  {{Hangup Kicks: Still Larger Recoils by Partial Spin/Orbit Alignment of
  Black-Hole Binaries}},}\ }\href {\doibase 10.1103/PhysRevLett.107.231102}
  {\bibfield  {journal} {\bibinfo  {journal} {Phys. Rev. Lett.}\ }\textbf
  {\bibinfo {volume} {107}},\ \bibinfo {pages} {231102} (\bibinfo {year}
  {2011})},\ \Eprint {http://arxiv.org/abs/1108.2009} {arXiv:1108.2009 [gr-qc]}
  \BibitemShut {NoStop}%
\bibitem [{\citenamefont {Lousto}\ and\ \citenamefont
  {Healy}(2019)}]{Lousto:2019lyf}%
  \BibitemOpen
  \bibfield  {author} {\bibinfo {author} {\bibfnamefont {Carlos~O.}\
  \bibnamefont {Lousto}}\ and\ \bibinfo {author} {\bibfnamefont {James}\
  \bibnamefont {Healy}},\ }\bibfield  {title} {\enquote {\bibinfo {title}
  {{Kicking gravitational wave detectors with recoiling black holes}},}\ }\href
  {\doibase 10.1103/PhysRevD.100.104039} {\bibfield  {journal} {\bibinfo
  {journal} {Phys. Rev.}\ }\textbf {\bibinfo {volume} {D100}},\ \bibinfo
  {pages} {104039} (\bibinfo {year} {2019})},\ \Eprint
  {http://arxiv.org/abs/1908.04382} {arXiv:1908.04382 [gr-qc]} \BibitemShut
  {NoStop}%
\bibitem [{\citenamefont {Sperhake}\ \emph {et~al.}(2020)\citenamefont
  {Sperhake}, \citenamefont {Rosca-Mead}, \citenamefont {Gerosa},\ and\
  \citenamefont {Berti}}]{Sperhake:2019wwo}%
  \BibitemOpen
  \bibfield  {author} {\bibinfo {author} {\bibfnamefont {U.}~\bibnamefont
  {Sperhake}}, \bibinfo {author} {\bibfnamefont {R.}~\bibnamefont
  {Rosca-Mead}}, \bibinfo {author} {\bibfnamefont {D.}~\bibnamefont {Gerosa}},
  \ and\ \bibinfo {author} {\bibfnamefont {E.}~\bibnamefont {Berti}},\
  }\bibfield  {title} {\enquote {\bibinfo {title} {{Amplification of superkicks
  in black-hole binaries through orbital eccentricity}},}\ }\href {\doibase
  10.1103/PhysRevD.101.024044} {\bibfield  {journal} {\bibinfo  {journal}
  {Phys. Rev.}\ }\textbf {\bibinfo {volume} {D101}},\ \bibinfo {pages} {024044}
  (\bibinfo {year} {2020})},\ \Eprint {http://arxiv.org/abs/1910.01598}
  {arXiv:1910.01598 [gr-qc]} \BibitemShut {NoStop}%
\bibitem [{\citenamefont {Merritt}\ \emph {et~al.}(2004)\citenamefont
  {Merritt}, \citenamefont {Milosavljevic}, \citenamefont {Favata},
  \citenamefont {Hughes},\ and\ \citenamefont {Holz}}]{Merritt:2004xa}%
  \BibitemOpen
  \bibfield  {author} {\bibinfo {author} {\bibfnamefont {David}\ \bibnamefont
  {Merritt}}, \bibinfo {author} {\bibfnamefont {Milos}\ \bibnamefont
  {Milosavljevic}}, \bibinfo {author} {\bibfnamefont {Marc}\ \bibnamefont
  {Favata}}, \bibinfo {author} {\bibfnamefont {Scott~A.}\ \bibnamefont
  {Hughes}}, \ and\ \bibinfo {author} {\bibfnamefont {Daniel~E.}\ \bibnamefont
  {Holz}},\ }\bibfield  {title} {\enquote {\bibinfo {title} {{Consequences of
  gravitational radiation recoil}},}\ }\href {\doibase 10.1086/421551}
  {\bibfield  {journal} {\bibinfo  {journal} {Astrophys. J.}\ }\textbf
  {\bibinfo {volume} {607}},\ \bibinfo {pages} {L9--L12} (\bibinfo {year}
  {2004})},\ \Eprint {http://arxiv.org/abs/astro-ph/0402057}
  {arXiv:astro-ph/0402057 [astro-ph]} \BibitemShut {NoStop}%
\bibitem [{\citenamefont {Komossa}\ \emph {et~al.}(2008)\citenamefont
  {Komossa}, \citenamefont {Zhou},\ and\ \citenamefont {Lu}}]{Komossa:2008qd}%
  \BibitemOpen
  \bibfield  {author} {\bibinfo {author} {\bibfnamefont {S.}~\bibnamefont
  {Komossa}}, \bibinfo {author} {\bibfnamefont {H.}~\bibnamefont {Zhou}}, \
  and\ \bibinfo {author} {\bibfnamefont {H.}~\bibnamefont {Lu}},\ }\bibfield
  {title} {\enquote {\bibinfo {title} {{A recoiling supermassive black hole in
  the quasar SDSSJ092712.65+294344.0?}}}\ }\href {\doibase 10.1086/588656}
  {\bibfield  {journal} {\bibinfo  {journal} {Astrophys. J.}\ }\textbf
  {\bibinfo {volume} {678}},\ \bibinfo {pages} {L81--L84} (\bibinfo {year}
  {2008})},\ \Eprint {http://arxiv.org/abs/0804.4585} {arXiv:0804.4585
  [astro-ph]} \BibitemShut {NoStop}%
\bibitem [{\citenamefont {Volonteri}\ \emph {et~al.}(2010)\citenamefont
  {Volonteri}, \citenamefont {Gültekin},\ and\ \citenamefont
  {Dotti}}]{Volonteri:2010mbh}%
  \BibitemOpen
  \bibfield  {author} {\bibinfo {author} {\bibfnamefont {Marta}\ \bibnamefont
  {Volonteri}}, \bibinfo {author} {\bibfnamefont {Kayhan}\ \bibnamefont
  {Gültekin}}, \ and\ \bibinfo {author} {\bibfnamefont {Massimo}\ \bibnamefont
  {Dotti}},\ }\bibfield  {title} {\enquote {\bibinfo {title} {{Gravitational
  recoil: effects on massive black hole occupation fraction over cosmic
  time}},}\ }\href {\doibase 10.1111/j.1365-2966.2010.16431.x} {\bibfield
  {journal} {\bibinfo  {journal} {Monthly Notices of the Royal Astronomical
  Society}\ }\textbf {\bibinfo {volume} {404}},\ \bibinfo {pages} {2143--2150}
  (\bibinfo {year} {2010})},\ \Eprint
  {http://arxiv.org/abs/http://oup.prod.sis.lan/mnras/article-pdf/404/4/2143/3803490/mnras0404-2143.pdf}
  {http://oup.prod.sis.lan/mnras/article-pdf/404/4/2143/3803490/mnras0404-2143.pdf}
  \BibitemShut {NoStop}%
\bibitem [{\citenamefont {Komossa}\ and\ \citenamefont
  {Merritt}(2008)}]{Komossa:2008as}%
  \BibitemOpen
  \bibfield  {author} {\bibinfo {author} {\bibfnamefont {S.}~\bibnamefont
  {Komossa}}\ and\ \bibinfo {author} {\bibfnamefont {David}\ \bibnamefont
  {Merritt}},\ }\bibfield  {title} {\enquote {\bibinfo {title} {{Gravitational
  Wave Recoil Oscillations of Black Holes: Implications for Unified Models of
  Active Galactic Nuclei}},}\ }\href {\doibase 10.1086/595883} {\bibfield
  {journal} {\bibinfo  {journal} {Astrophys. J.}\ }\textbf {\bibinfo {volume}
  {689}},\ \bibinfo {pages} {L89} (\bibinfo {year} {2008})},\ \Eprint
  {http://arxiv.org/abs/0811.1037} {arXiv:0811.1037 [astro-ph]} \BibitemShut
  {NoStop}%
\bibitem [{\citenamefont {Gerosa}\ and\ \citenamefont
  {Sesana}(2015)}]{Gerosa:2014gja}%
  \BibitemOpen
  \bibfield  {author} {\bibinfo {author} {\bibfnamefont {Davide}\ \bibnamefont
  {Gerosa}}\ and\ \bibinfo {author} {\bibfnamefont {Alberto}\ \bibnamefont
  {Sesana}},\ }\bibfield  {title} {\enquote {\bibinfo {title} {{Missing black
  holes in brightest cluster galaxies as evidence for the occurrence of
  superkicks in nature}},}\ }\href {\doibase 10.1093/mnras/stu2049} {\bibfield
  {journal} {\bibinfo  {journal} {Mon. Not. Roy. Astron. Soc.}\ }\textbf
  {\bibinfo {volume} {446}},\ \bibinfo {pages} {38--55} (\bibinfo {year}
  {2015})},\ \Eprint {http://arxiv.org/abs/1405.2072} {arXiv:1405.2072
  [astro-ph.GA]} \BibitemShut {NoStop}%
\bibitem [{\citenamefont {sesana}(2007)}]{sesana:2007zk}%
  \BibitemOpen
  \bibfield  {author} {\bibinfo {author} {\bibfnamefont {A.}~\bibnamefont
  {sesana}},\ }\bibfield  {title} {\enquote {\bibinfo {title} {{Extreme
  recoils: impact on the detection of gravitational waves from massive black
  hole binaries}},}\ }\href {\doibase 10.1111/j.1745-3933.2007.00375.x}
  {\bibfield  {journal} {\bibinfo  {journal} {Mon. Not. Roy. Astron. Soc.}\
  }\textbf {\bibinfo {volume} {382}},\ \bibinfo {pages} {6} (\bibinfo {year}
  {2007})},\ \Eprint {http://arxiv.org/abs/0707.4677} {arXiv:0707.4677
  [astro-ph]} \BibitemShut {NoStop}%
\bibitem [{\citenamefont {Amaro-Seoane}\ \emph {et~al.}(2017)\citenamefont
  {Amaro-Seoane} \emph {et~al.}}]{amaroseoane2017laser}%
  \BibitemOpen
  \bibfield  {author} {\bibinfo {author} {\bibfnamefont {Pau}\ \bibnamefont
  {Amaro-Seoane}} \emph {et~al.},\ }\href@noop {} {\enquote {\bibinfo {title}
  {Laser interferometer space antenna},}\ } (\bibinfo {year} {2017}),\ \Eprint
  {http://arxiv.org/abs/1702.00786} {arXiv:1702.00786 [astro-ph.IM]}
  \BibitemShut {NoStop}%
\bibitem [{\citenamefont {Yang}\ \emph {et~al.}(2019)\citenamefont {Yang} \emph
  {et~al.}}]{Yang:2019cbr}%
  \BibitemOpen
  \bibfield  {author} {\bibinfo {author} {\bibfnamefont {Yang}\ \bibnamefont
  {Yang}} \emph {et~al.},\ }\bibfield  {title} {\enquote {\bibinfo {title}
  {{Hierarchical Black Hole Mergers in Active Galactic Nuclei}},}\ }\href
  {\doibase 10.1103/PhysRevLett.123.181101} {\bibfield  {journal} {\bibinfo
  {journal} {Phys. Rev. Lett.}\ }\textbf {\bibinfo {volume} {123}},\ \bibinfo
  {pages} {181101} (\bibinfo {year} {2019})},\ \Eprint
  {http://arxiv.org/abs/1906.09281} {arXiv:1906.09281 [astro-ph.HE]}
  \BibitemShut {NoStop}%
\bibitem [{\citenamefont {Gayathri}\ \emph {et~al.}(2020)\citenamefont
  {Gayathri}, \citenamefont {Bartos}, \citenamefont {Haiman}, \citenamefont
  {Klimenko}, \citenamefont {Kocsis}, \citenamefont {Marka},\ and\
  \citenamefont {Yang}}]{Gayathri:2019kop}%
  \BibitemOpen
  \bibfield  {author} {\bibinfo {author} {\bibfnamefont {V.}~\bibnamefont
  {Gayathri}}, \bibinfo {author} {\bibfnamefont {I.}~\bibnamefont {Bartos}},
  \bibinfo {author} {\bibfnamefont {Z.}~\bibnamefont {Haiman}}, \bibinfo
  {author} {\bibfnamefont {S.}~\bibnamefont {Klimenko}}, \bibinfo {author}
  {\bibfnamefont {B.}~\bibnamefont {Kocsis}}, \bibinfo {author} {\bibfnamefont
  {S.}~\bibnamefont {Marka}}, \ and\ \bibinfo {author} {\bibfnamefont
  {Y.}~\bibnamefont {Yang}},\ }\bibfield  {title} {\enquote {\bibinfo {title}
  {{GW170817A as a Hierarchical Black Hole Merger}},}\ }\href {\doibase
  10.3847/2041-8213/ab745d} {\bibfield  {journal} {\bibinfo  {journal}
  {Astrophys. J.}\ }\textbf {\bibinfo {volume} {890}},\ \bibinfo {pages} {L20}
  (\bibinfo {year} {2020})},\ \Eprint {http://arxiv.org/abs/1911.11142}
  {arXiv:1911.11142 [astro-ph.HE]} \BibitemShut {NoStop}%
\bibitem [{\citenamefont {Gerosa}\ and\ \citenamefont
  {Berti}(2019)}]{Gerosa:2019zmo}%
  \BibitemOpen
  \bibfield  {author} {\bibinfo {author} {\bibfnamefont {Davide}\ \bibnamefont
  {Gerosa}}\ and\ \bibinfo {author} {\bibfnamefont {Emanuele}\ \bibnamefont
  {Berti}},\ }\bibfield  {title} {\enquote {\bibinfo {title} {{Escape speed of
  stellar clusters from multiple-generation black-hole mergers in the upper
  mass gap}},}\ }\href {\doibase 10.1103/PhysRevD.100.041301} {\bibfield
  {journal} {\bibinfo  {journal} {Phys. Rev.}\ }\textbf {\bibinfo {volume}
  {D100}},\ \bibinfo {pages} {041301} (\bibinfo {year} {2019})},\ \Eprint
  {http://arxiv.org/abs/1906.05295} {arXiv:1906.05295 [astro-ph.HE]}
  \BibitemShut {NoStop}%
\bibitem [{\citenamefont {Mangiagli}\ \emph {et~al.}(2019)\citenamefont
  {Mangiagli}, \citenamefont {Bonetti}, \citenamefont {Sesana},\ and\
  \citenamefont {Colpi}}]{Mangiagli:2019sxg}%
  \BibitemOpen
  \bibfield  {author} {\bibinfo {author} {\bibfnamefont {Alberto}\ \bibnamefont
  {Mangiagli}}, \bibinfo {author} {\bibfnamefont {Matteo}\ \bibnamefont
  {Bonetti}}, \bibinfo {author} {\bibfnamefont {Alberto}\ \bibnamefont
  {Sesana}}, \ and\ \bibinfo {author} {\bibfnamefont {Monica}\ \bibnamefont
  {Colpi}},\ }\bibfield  {title} {\enquote {\bibinfo {title} {{Merger rate of
  stellar black hole binaries above the pair instability mass gap}},}\ }\href
  {\doibase 10.3847/2041-8213/ab3f33} {\bibfield  {journal} {\bibinfo
  {journal} {Astrophys. J.}\ }\textbf {\bibinfo {volume} {883}},\ \bibinfo
  {pages} {L27} (\bibinfo {year} {2019})},\ \Eprint
  {http://arxiv.org/abs/1907.12562} {arXiv:1907.12562 [astro-ph.HE]}
  \BibitemShut {NoStop}%
\bibitem [{\citenamefont {Di~Carlo}\ \emph {et~al.}(2019)\citenamefont
  {Di~Carlo}, \citenamefont {Mapelli}, \citenamefont {Bouffanais},
  \citenamefont {Giacobbo}, \citenamefont {Bressan}, \citenamefont {Spera},\
  and\ \citenamefont {Haardt}}]{DiCarlo:2019fcq}%
  \BibitemOpen
  \bibfield  {author} {\bibinfo {author} {\bibfnamefont {Ugo~N.}\ \bibnamefont
  {Di~Carlo}}, \bibinfo {author} {\bibfnamefont {Michela}\ \bibnamefont
  {Mapelli}}, \bibinfo {author} {\bibfnamefont {Yann}\ \bibnamefont
  {Bouffanais}}, \bibinfo {author} {\bibfnamefont {Nicola}\ \bibnamefont
  {Giacobbo}}, \bibinfo {author} {\bibfnamefont {Sandro}\ \bibnamefont
  {Bressan}}, \bibinfo {author} {\bibfnamefont {Mario}\ \bibnamefont {Spera}},
  \ and\ \bibinfo {author} {\bibfnamefont {Francesco}\ \bibnamefont {Haardt}},\
  }\bibfield  {title} {\enquote {\bibinfo {title} {{Binary black holes in the
  pair-instability mass gap}},}\ }\href@noop {} {\  (\bibinfo {year} {2019})},\
  \Eprint {http://arxiv.org/abs/1911.01434} {arXiv:1911.01434 [astro-ph.HE]}
  \BibitemShut {NoStop}%
\bibitem [{\citenamefont {Rodriguez}\ \emph {et~al.}(2019)\citenamefont
  {Rodriguez}, \citenamefont {Zevin}, \citenamefont {Amaro-Seoane},
  \citenamefont {Chatterjee}, \citenamefont {Kremer}, \citenamefont {Rasio},\
  and\ \citenamefont {Ye}}]{Rodriguez:2019huv}%
  \BibitemOpen
  \bibfield  {author} {\bibinfo {author} {\bibfnamefont {Carl~L.}\ \bibnamefont
  {Rodriguez}}, \bibinfo {author} {\bibfnamefont {Michael}\ \bibnamefont
  {Zevin}}, \bibinfo {author} {\bibfnamefont {Pau}\ \bibnamefont
  {Amaro-Seoane}}, \bibinfo {author} {\bibfnamefont {Sourav}\ \bibnamefont
  {Chatterjee}}, \bibinfo {author} {\bibfnamefont {Kyle}\ \bibnamefont
  {Kremer}}, \bibinfo {author} {\bibfnamefont {Frederic~A.}\ \bibnamefont
  {Rasio}}, \ and\ \bibinfo {author} {\bibfnamefont {Claire~S.}\ \bibnamefont
  {Ye}},\ }\bibfield  {title} {\enquote {\bibinfo {title} {{Black holes: The
  next generation—repeated mergers in dense star clusters and their
  gravitational-wave properties}},}\ }\href {\doibase
  10.1103/PhysRevD.100.043027} {\bibfield  {journal} {\bibinfo  {journal}
  {Phys. Rev.}\ }\textbf {\bibinfo {volume} {D100}},\ \bibinfo {pages} {043027}
  (\bibinfo {year} {2019})},\ \Eprint {http://arxiv.org/abs/1906.10260}
  {arXiv:1906.10260 [astro-ph.HE]} \BibitemShut {NoStop}%
\bibitem [{\citenamefont {Doctor}\ \emph {et~al.}(2019)\citenamefont {Doctor},
  \citenamefont {Wysocki}, \citenamefont {O'Shaughnessy}, \citenamefont
  {Holz},\ and\ \citenamefont {Farr}}]{Doctor:2019ruh}%
  \BibitemOpen
  \bibfield  {author} {\bibinfo {author} {\bibfnamefont {Zoheyr}\ \bibnamefont
  {Doctor}}, \bibinfo {author} {\bibfnamefont {Daniel}\ \bibnamefont
  {Wysocki}}, \bibinfo {author} {\bibfnamefont {Richard}\ \bibnamefont
  {O'Shaughnessy}}, \bibinfo {author} {\bibfnamefont {Daniel~E.}\ \bibnamefont
  {Holz}}, \ and\ \bibinfo {author} {\bibfnamefont {Ben}\ \bibnamefont
  {Farr}},\ }\bibfield  {title} {\enquote {\bibinfo {title} {{Black Hole
  Coagulation: Modeling Hierarchical Mergers in Black Hole Populations}},}\
  }\href@noop {} {\  (\bibinfo {year} {2019})},\ \Eprint
  {http://arxiv.org/abs/1911.04424} {arXiv:1911.04424 [astro-ph.HE]}
  \BibitemShut {NoStop}%
\bibitem [{\citenamefont {Farmer}\ \emph {et~al.}(2019)\citenamefont {Farmer},
  \citenamefont {Renzo}, \citenamefont {de~Mink}, \citenamefont {Marchant},\
  and\ \citenamefont {Justham}}]{Farmer:2019jed}%
  \BibitemOpen
  \bibfield  {author} {\bibinfo {author} {\bibfnamefont {R.}~\bibnamefont
  {Farmer}}, \bibinfo {author} {\bibfnamefont {M.}~\bibnamefont {Renzo}},
  \bibinfo {author} {\bibfnamefont {S.~E.}\ \bibnamefont {de~Mink}}, \bibinfo
  {author} {\bibfnamefont {P.}~\bibnamefont {Marchant}}, \ and\ \bibinfo
  {author} {\bibfnamefont {S.}~\bibnamefont {Justham}},\ }\bibfield  {title}
  {\enquote {\bibinfo {title} {{Mind the gap: The location of the lower edge of
  the pair instability supernovae black hole mass gap}},}\ }\href@noop {} {\
  (\bibinfo {year} {2019})},\ \Eprint {http://arxiv.org/abs/1910.12874}
  {arXiv:1910.12874 [astro-ph.SR]} \BibitemShut {NoStop}%
\bibitem [{\citenamefont {Abbott}\ \emph
  {et~al.}(2019{\natexlab{b}})\citenamefont {Abbott} \emph
  {et~al.}}]{LIGOScientific:2018mvr}%
  \BibitemOpen
  \bibfield  {author} {\bibinfo {author} {\bibfnamefont {B.~P.}\ \bibnamefont
  {Abbott}} \emph {et~al.} (\bibinfo {collaboration} {LIGO Scientific,
  Virgo}),\ }\bibfield  {title} {\enquote {\bibinfo {title} {{GWTC-1: A
  Gravitational-Wave Transient Catalog of Compact Binary Mergers Observed by
  LIGO and Virgo during the First and Second Observing Runs}},}\ }\href
  {\doibase 10.1103/PhysRevX.9.031040} {\bibfield  {journal} {\bibinfo
  {journal} {Phys. Rev.}\ }\textbf {\bibinfo {volume} {X9}},\ \bibinfo {pages}
  {031040} (\bibinfo {year} {2019}{\natexlab{b}})},\ \Eprint
  {http://arxiv.org/abs/1811.12907} {arXiv:1811.12907 [astro-ph.HE]}
  \BibitemShut {NoStop}%
\bibitem [{\citenamefont {Chatziioannou}\ \emph {et~al.}(2019)\citenamefont
  {Chatziioannou} \emph {et~al.}}]{Chatziioannou:2019dsz}%
  \BibitemOpen
  \bibfield  {author} {\bibinfo {author} {\bibfnamefont {Katerina}\
  \bibnamefont {Chatziioannou}} \emph {et~al.},\ }\bibfield  {title} {\enquote
  {\bibinfo {title} {{On the properties of the massive binary black hole merger
  GW170729}},}\ }\href {\doibase 10.1103/PhysRevD.100.104015} {\bibfield
  {journal} {\bibinfo  {journal} {Phys. Rev.}\ }\textbf {\bibinfo {volume}
  {D100}},\ \bibinfo {pages} {104015} (\bibinfo {year} {2019})},\ \Eprint
  {http://arxiv.org/abs/1903.06742} {arXiv:1903.06742 [gr-qc]} \BibitemShut
  {NoStop}%
\bibitem [{\citenamefont {Woosley}(2017)}]{Woosley:2016hmi}%
  \BibitemOpen
  \bibfield  {author} {\bibinfo {author} {\bibfnamefont {S.~E.}\ \bibnamefont
  {Woosley}},\ }\bibfield  {title} {\enquote {\bibinfo {title} {{Pulsational
  Pair-Instability Supernovae}},}\ }\href {\doibase
  10.3847/1538-4357/836/2/244} {\bibfield  {journal} {\bibinfo  {journal}
  {Astrophys. J.}\ }\textbf {\bibinfo {volume} {836}},\ \bibinfo {pages} {244}
  (\bibinfo {year} {2017})},\ \Eprint {http://arxiv.org/abs/1608.08939}
  {arXiv:1608.08939 [astro-ph.HE]} \BibitemShut {NoStop}%
\bibitem [{\citenamefont {Marchant}\ \emph {et~al.}(2018)\citenamefont
  {Marchant}, \citenamefont {Renzo}, \citenamefont {Farmer}, \citenamefont
  {Pappas}, \citenamefont {Taam}, \citenamefont {de~Mink},\ and\ \citenamefont
  {Kalogera}}]{Marchant:2018kun}%
  \BibitemOpen
  \bibfield  {author} {\bibinfo {author} {\bibfnamefont {Pablo}\ \bibnamefont
  {Marchant}}, \bibinfo {author} {\bibfnamefont {Mathieu}\ \bibnamefont
  {Renzo}}, \bibinfo {author} {\bibfnamefont {Robert}\ \bibnamefont {Farmer}},
  \bibinfo {author} {\bibfnamefont {Kaliroe M.~W.}\ \bibnamefont {Pappas}},
  \bibinfo {author} {\bibfnamefont {Ronald~E.}\ \bibnamefont {Taam}}, \bibinfo
  {author} {\bibfnamefont {Selma}\ \bibnamefont {de~Mink}}, \ and\ \bibinfo
  {author} {\bibfnamefont {Vassiliki}\ \bibnamefont {Kalogera}},\ }\bibfield
  {title} {\enquote {\bibinfo {title} {{Pulsational pair-instability supernovae
  in very close binaries}},}\ }\href {\doibase 10.3847/1538-4357/ab3426} {\
  (\bibinfo {year} {2018}),\ 10.3847/1538-4357/ab3426},\ \Eprint
  {http://arxiv.org/abs/1810.13412} {arXiv:1810.13412 [astro-ph.HE]}
  \BibitemShut {NoStop}%
\bibitem [{\citenamefont {Varma}\ \emph
  {et~al.}(2019{\natexlab{a}})\citenamefont {Varma}, \citenamefont {Field},
  \citenamefont {Scheel}, \citenamefont {Blackman}, \citenamefont {Gerosa},
  \citenamefont {Stein}, \citenamefont {Kidder},\ and\ \citenamefont
  {Pfeiffer}}]{Varma:2019csw}%
  \BibitemOpen
  \bibfield  {author} {\bibinfo {author} {\bibfnamefont {Vijay}\ \bibnamefont
  {Varma}}, \bibinfo {author} {\bibfnamefont {Scott~E.}\ \bibnamefont {Field}},
  \bibinfo {author} {\bibfnamefont {Mark~A.}\ \bibnamefont {Scheel}}, \bibinfo
  {author} {\bibfnamefont {Jonathan}\ \bibnamefont {Blackman}}, \bibinfo
  {author} {\bibfnamefont {Davide}\ \bibnamefont {Gerosa}}, \bibinfo {author}
  {\bibfnamefont {Leo~C.}\ \bibnamefont {Stein}}, \bibinfo {author}
  {\bibfnamefont {Lawrence~E.}\ \bibnamefont {Kidder}}, \ and\ \bibinfo
  {author} {\bibfnamefont {Harald~P.}\ \bibnamefont {Pfeiffer}},\ }\bibfield
  {title} {\enquote {\bibinfo {title} {{Surrogate models for precessing binary
  black hole simulations with unequal masses}},}\ }\href {\doibase
  10.1103/PhysRevResearch.1.033015} {\bibfield  {journal} {\bibinfo  {journal}
  {Phys. Rev. Research.}\ }\textbf {\bibinfo {volume} {1}},\ \bibinfo {pages}
  {033015} (\bibinfo {year} {2019}{\natexlab{a}})},\ \Eprint
  {http://arxiv.org/abs/1905.09300} {arXiv:1905.09300 [gr-qc]} \BibitemShut
  {NoStop}%
\bibitem [{\citenamefont {Blackman}\ \emph {et~al.}(2017)\citenamefont
  {Blackman}, \citenamefont {Field}, \citenamefont {Scheel}, \citenamefont
  {Galley}, \citenamefont {Ott}, \citenamefont {Boyle}, \citenamefont {Kidder},
  \citenamefont {Pfeiffer},\ and\ \citenamefont
  {Szilágyi}}]{Blackman:2017pcm}%
  \BibitemOpen
  \bibfield  {author} {\bibinfo {author} {\bibfnamefont {Jonathan}\
  \bibnamefont {Blackman}}, \bibinfo {author} {\bibfnamefont {Scott~E.}\
  \bibnamefont {Field}}, \bibinfo {author} {\bibfnamefont {Mark~A.}\
  \bibnamefont {Scheel}}, \bibinfo {author} {\bibfnamefont {Chad~R.}\
  \bibnamefont {Galley}}, \bibinfo {author} {\bibfnamefont {Christian~D.}\
  \bibnamefont {Ott}}, \bibinfo {author} {\bibfnamefont {Michael}\ \bibnamefont
  {Boyle}}, \bibinfo {author} {\bibfnamefont {Lawrence~E.}\ \bibnamefont
  {Kidder}}, \bibinfo {author} {\bibfnamefont {Harald~P.}\ \bibnamefont
  {Pfeiffer}}, \ and\ \bibinfo {author} {\bibfnamefont {Béla}\ \bibnamefont
  {Szilágyi}},\ }\bibfield  {title} {\enquote {\bibinfo {title} {{Numerical
  relativity waveform surrogate model for generically precessing binary black
  hole mergers}},}\ }\href {\doibase 10.1103/PhysRevD.96.024058} {\bibfield
  {journal} {\bibinfo  {journal} {Phys. Rev.}\ }\textbf {\bibinfo {volume}
  {D96}},\ \bibinfo {pages} {024058} (\bibinfo {year} {2017})},\ \Eprint
  {http://arxiv.org/abs/1705.07089} {arXiv:1705.07089 [gr-qc]} \BibitemShut
  {NoStop}%
\bibitem [{\citenamefont {Varma}\ \emph
  {et~al.}(2019{\natexlab{b}})\citenamefont {Varma}, \citenamefont {Gerosa},
  \citenamefont {Stein}, \citenamefont {Hébert},\ and\ \citenamefont
  {Zhang}}]{Varma:2018aht}%
  \BibitemOpen
  \bibfield  {author} {\bibinfo {author} {\bibfnamefont {Vijay}\ \bibnamefont
  {Varma}}, \bibinfo {author} {\bibfnamefont {Davide}\ \bibnamefont {Gerosa}},
  \bibinfo {author} {\bibfnamefont {Leo~C.}\ \bibnamefont {Stein}}, \bibinfo
  {author} {\bibfnamefont {François}\ \bibnamefont {Hébert}}, \ and\ \bibinfo
  {author} {\bibfnamefont {Hao}\ \bibnamefont {Zhang}},\ }\bibfield  {title}
  {\enquote {\bibinfo {title} {{High-accuracy mass, spin, and recoil
  predictions of generic black-hole merger remnants}},}\ }\href {\doibase
  10.1103/PhysRevLett.122.011101} {\bibfield  {journal} {\bibinfo  {journal}
  {Phys. Rev. Lett.}\ }\textbf {\bibinfo {volume} {122}},\ \bibinfo {pages}
  {011101} (\bibinfo {year} {2019}{\natexlab{b}})},\ \Eprint
  {http://arxiv.org/abs/1809.09125} {arXiv:1809.09125 [gr-qc]} \BibitemShut
  {NoStop}%
\bibitem [{\citenamefont {{LIGO Scientific Collaboration and Virgo
  Collaboration}}(2018)}]{GWOSC:GWTC}%
  \BibitemOpen
  \bibfield  {author} {\bibinfo {author} {\bibnamefont {{LIGO Scientific
  Collaboration and Virgo Collaboration}}},\ }\href@noop {} {\enquote {\bibinfo
  {title} {{GWTC-1}},}\ }\bibinfo {howpublished}
  {\href{https://doi.org/10.7935/82H3-HH23}{https://doi.org/10.7935/82H3-HH23}}
  (\bibinfo {year} {2018})\BibitemShut {NoStop}%
\bibitem [{\citenamefont {{LIGO Scientific
  Collaboration}}(2018{\natexlab{a}})}]{aLIGODesignNoiseCurve}%
  \BibitemOpen
  \bibfield  {author} {\bibinfo {author} {\bibnamefont {{LIGO Scientific
  Collaboration}}},\ }\href@noop {} {\emph {\bibinfo {title} {Updated Advanced
  LIGO sensitivity design curve}}},\ \bibinfo {type} {Tech. Rep.}\ (\bibinfo
  {year} {2018})\ \bibinfo {note}
  {\url{https://dcc.ligo.org/LIGO-T1800044/public}}\BibitemShut {NoStop}%
\bibitem [{\citenamefont {Manzotti}\ and\ \citenamefont
  {Dietz}(2012)}]{Manzotti:2012uw}%
  \BibitemOpen
  \bibfield  {author} {\bibinfo {author} {\bibfnamefont {Alessandro}\
  \bibnamefont {Manzotti}}\ and\ \bibinfo {author} {\bibfnamefont {Alexander}\
  \bibnamefont {Dietz}},\ }\bibfield  {title} {\enquote {\bibinfo {title}
  {{Prospects for early localization of gravitational-wave signals from compact
  binary coalescences with advanced detectors}},}\ }\href@noop {} {\  (\bibinfo
  {year} {2012})},\ \Eprint {http://arxiv.org/abs/1202.4031} {arXiv:1202.4031
  [gr-qc]} \BibitemShut {NoStop}%
\bibitem [{\citenamefont {Abbott}\ \emph {et~al.}(2018)\citenamefont {Abbott}
  \emph {et~al.}}]{Aasi:2013wya}%
  \BibitemOpen
  \bibfield  {author} {\bibinfo {author} {\bibfnamefont {B.~P.}\ \bibnamefont
  {Abbott}} \emph {et~al.} (\bibinfo {collaboration} {KAGRA, LIGO Scientific,
  VIRGO}),\ }\bibfield  {title} {\enquote {\bibinfo {title} {{Prospects for
  Observing and Localizing Gravitational-Wave Transients with Advanced LIGO,
  Advanced Virgo and KAGRA}},}\ }\href {\doibase 10.1007/s41114-018-0012-9,
  10.1007/lrr-2016-1} {\bibfield  {journal} {\bibinfo  {journal} {Living Rev.
  Rel.}\ }\textbf {\bibinfo {volume} {21}},\ \bibinfo {pages} {3} (\bibinfo
  {year} {2018})},\ \Eprint {http://arxiv.org/abs/1304.0670} {arXiv:1304.0670
  [gr-qc]} \BibitemShut {NoStop}%
\bibitem [{\citenamefont {Veitch}\ \emph {et~al.}(2015)\citenamefont {Veitch}
  \emph {et~al.}}]{Veitch:2014wba}%
  \BibitemOpen
  \bibfield  {author} {\bibinfo {author} {\bibfnamefont {J.}~\bibnamefont
  {Veitch}} \emph {et~al.},\ }\bibfield  {title} {\enquote {\bibinfo {title}
  {{Robust parameter estimation for compact binaries with ground-based
  gravitational-wave observations using the LALInference software library}},}\
  }\href {\doibase 10.1103/PhysRevD.91.042003} {\bibfield  {journal} {\bibinfo
  {journal} {Phys. Rev. D}\ }\textbf {\bibinfo {volume} {91}},\ \bibinfo
  {pages} {042003} (\bibinfo {year} {2015})},\ \Eprint
  {http://arxiv.org/abs/1409.7215} {arXiv:1409.7215 [gr-qc]} \BibitemShut
  {NoStop}%
\bibitem [{\citenamefont {{LIGO Scientific
  Collaboration}}(2018{\natexlab{b}})}]{lalsuite}%
  \BibitemOpen
  \bibfield  {author} {\bibinfo {author} {\bibnamefont {{LIGO Scientific
  Collaboration}}},\ }\href {\doibase 10.7935/GT1W-FZ16} {\enquote {\bibinfo
  {title} {{LIGO} {A}lgorithm {L}ibrary - {LALS}uite},}\ }\bibinfo
  {howpublished} {free software (GPL)} (\bibinfo {year}
  {2018}{\natexlab{b}})\BibitemShut {NoStop}%
\bibitem [{\citenamefont {Schmidt}\ \emph {et~al.}(2017)\citenamefont
  {Schmidt}, \citenamefont {Harry},\ and\ \citenamefont
  {Pfeiffer}}]{Schmidt:2017btt}%
  \BibitemOpen
  \bibfield  {author} {\bibinfo {author} {\bibfnamefont {Patricia}\
  \bibnamefont {Schmidt}}, \bibinfo {author} {\bibfnamefont {Ian~W.}\
  \bibnamefont {Harry}}, \ and\ \bibinfo {author} {\bibfnamefont {Harald~P.}\
  \bibnamefont {Pfeiffer}},\ }\href@noop {} {\enquote {\bibinfo {title}
  {{Numerical Relativity Injection Infrastructure}},}\ } (\bibinfo {year}
  {2017}),\ \Eprint {http://arxiv.org/abs/1703.01076} {arXiv:1703.01076
  [gr-qc]} \BibitemShut {NoStop}%
\bibitem [{\citenamefont {{Rasmussen}}\ and\ \citenamefont
  {{Williams}}(2006)}]{Rasmussen_Williams_GPRbook}%
  \BibitemOpen
  \bibfield  {author} {\bibinfo {author} {\bibfnamefont {C.~E.}\ \bibnamefont
  {{Rasmussen}}}\ and\ \bibinfo {author} {\bibfnamefont {C.~K.~I.}\
  \bibnamefont {{Williams}}},\ }\href@noop {} {\emph {\bibinfo {title}
  {Gaussian Processes for Machine Learning, by C.E.~Rasmussen and
  C.K.I.~Williams.~ISBN-13 978-0-262-18253-9}}}\ (\bibinfo {year}
  {2006})\BibitemShut {NoStop}%
\bibitem [{kic({\natexlab{a}})}]{kickpapersupplement}%
  \BibitemOpen
  \href@noop {} {}\bibinfo {howpublished} {See Supplemental Material
  \hyperlink{page.9}{here}, for details about the Doppler shifted remnant mass,
  a study of biases in the kick measurement, and priors used for the component
  parameters. This further includes Refs.~\cite{Krolak1987, Barausse:2012qz,
  Hofmann:2016yih, Jimenez-Forteza:2016oae, Healy:2016lce, Healy:2014yta,
  Vishveshwara:1970, Press:1971ApJ, Teukolsky:1973ApJ,
  Chandrasekhar_Detweiler:1975, Gair:2015nga, Dreyer:2003bv, Gossan:2011ha,
  Meidam:2014jpa}.} ({\natexlab{a}})\BibitemShut {NoStop}%
\bibitem [{\citenamefont {Krolak}\ and\ \citenamefont
  {Schutz}(1987)}]{Krolak1987}%
  \BibitemOpen
  \bibfield  {author} {\bibinfo {author} {\bibfnamefont {A}~\bibnamefont
  {Krolak}}\ and\ \bibinfo {author} {\bibfnamefont {Bernard~F}\ \bibnamefont
  {Schutz}},\ }\bibfield  {title} {\enquote {\bibinfo {title} {{Coalescing
  binaries—Probe of the universe}},}\ }\href {\doibase 10.1007/BF00759095}
  {\bibfield  {journal} {\bibinfo  {journal} {General Relativity and
  Gravitation}\ }\textbf {\bibinfo {volume} {19}},\ \bibinfo {pages}
  {1163--1171} (\bibinfo {year} {1987})}\BibitemShut {NoStop}%
\bibitem [{\citenamefont {Barausse}\ \emph {et~al.}(2012)\citenamefont
  {Barausse}, \citenamefont {Morozova},\ and\ \citenamefont
  {Rezzolla}}]{Barausse:2012qz}%
  \BibitemOpen
  \bibfield  {author} {\bibinfo {author} {\bibfnamefont {Enrico}\ \bibnamefont
  {Barausse}}, \bibinfo {author} {\bibfnamefont {Viktoriya}\ \bibnamefont
  {Morozova}}, \ and\ \bibinfo {author} {\bibfnamefont {Luciano}\ \bibnamefont
  {Rezzolla}},\ }\bibfield  {title} {\enquote {\bibinfo {title} {{On the mass
  radiated by coalescing black-hole binaries}},}\ }\href {\doibase
  10.1088/0004-637X/758/1/63} {\bibfield  {journal} {\bibinfo  {journal}
  {Astrophys. J.}\ }\textbf {\bibinfo {volume} {758}},\ \bibinfo {pages} {63}
  (\bibinfo {year} {2012})},\ \bibinfo {note} {[Erratum: Astrophys.
  J.786,76(2014)]},\ \Eprint {http://arxiv.org/abs/1206.3803} {arXiv:1206.3803
  [gr-qc]} \BibitemShut {NoStop}%
\bibitem [{\citenamefont {Hofmann}\ \emph {et~al.}(2016)\citenamefont
  {Hofmann}, \citenamefont {Barausse},\ and\ \citenamefont
  {Rezzolla}}]{Hofmann:2016yih}%
  \BibitemOpen
  \bibfield  {author} {\bibinfo {author} {\bibfnamefont {Fabian}\ \bibnamefont
  {Hofmann}}, \bibinfo {author} {\bibfnamefont {Enrico}\ \bibnamefont
  {Barausse}}, \ and\ \bibinfo {author} {\bibfnamefont {Luciano}\ \bibnamefont
  {Rezzolla}},\ }\bibfield  {title} {\enquote {\bibinfo {title} {{The final
  spin from binary black holes in quasi-circular orbits}},}\ }\href {\doibase
  10.3847/2041-8205/825/2/L19} {\bibfield  {journal} {\bibinfo  {journal}
  {Astrophys. J.}\ }\textbf {\bibinfo {volume} {825}},\ \bibinfo {pages} {L19}
  (\bibinfo {year} {2016})},\ \Eprint {http://arxiv.org/abs/1605.01938}
  {arXiv:1605.01938 [gr-qc]} \BibitemShut {NoStop}%
\bibitem [{\citenamefont {Jiménez-Forteza}\ \emph {et~al.}(2017)\citenamefont
  {Jiménez-Forteza}, \citenamefont {Keitel}, \citenamefont {Husa},
  \citenamefont {Hannam}, \citenamefont {Khan},\ and\ \citenamefont
  {Pürrer}}]{Jimenez-Forteza:2016oae}%
  \BibitemOpen
  \bibfield  {author} {\bibinfo {author} {\bibfnamefont {Xisco}\ \bibnamefont
  {Jiménez-Forteza}}, \bibinfo {author} {\bibfnamefont {David}\ \bibnamefont
  {Keitel}}, \bibinfo {author} {\bibfnamefont {Sascha}\ \bibnamefont {Husa}},
  \bibinfo {author} {\bibfnamefont {Mark}\ \bibnamefont {Hannam}}, \bibinfo
  {author} {\bibfnamefont {Sebastian}\ \bibnamefont {Khan}}, \ and\ \bibinfo
  {author} {\bibfnamefont {Michael}\ \bibnamefont {Pürrer}},\ }\bibfield
  {title} {\enquote {\bibinfo {title} {{Hierarchical data-driven approach to
  fitting numerical relativity data for nonprecessing binary black holes with
  an application to final spin and radiated energy}},}\ }\href {\doibase
  10.1103/PhysRevD.95.064024} {\bibfield  {journal} {\bibinfo  {journal} {Phys.
  Rev.}\ }\textbf {\bibinfo {volume} {D95}},\ \bibinfo {pages} {064024}
  (\bibinfo {year} {2017})},\ \Eprint {http://arxiv.org/abs/1611.00332}
  {arXiv:1611.00332 [gr-qc]} \BibitemShut {NoStop}%
\bibitem [{\citenamefont {Healy}\ and\ \citenamefont
  {Lousto}(2017)}]{Healy:2016lce}%
  \BibitemOpen
  \bibfield  {author} {\bibinfo {author} {\bibfnamefont {James}\ \bibnamefont
  {Healy}}\ and\ \bibinfo {author} {\bibfnamefont {Carlos~O.}\ \bibnamefont
  {Lousto}},\ }\bibfield  {title} {\enquote {\bibinfo {title} {{Remnant of
  binary black-hole mergers: New simulations and peak luminosity studies}},}\
  }\href {\doibase 10.1103/PhysRevD.95.024037} {\bibfield  {journal} {\bibinfo
  {journal} {Phys. Rev.}\ }\textbf {\bibinfo {volume} {D95}},\ \bibinfo {pages}
  {024037} (\bibinfo {year} {2017})},\ \Eprint
  {http://arxiv.org/abs/1610.09713} {arXiv:1610.09713 [gr-qc]} \BibitemShut
  {NoStop}%
\bibitem [{\citenamefont {Healy}\ \emph {et~al.}(2014)\citenamefont {Healy},
  \citenamefont {Lousto},\ and\ \citenamefont {Zlochower}}]{Healy:2014yta}%
  \BibitemOpen
  \bibfield  {author} {\bibinfo {author} {\bibfnamefont {James}\ \bibnamefont
  {Healy}}, \bibinfo {author} {\bibfnamefont {Carlos~O.}\ \bibnamefont
  {Lousto}}, \ and\ \bibinfo {author} {\bibfnamefont {Yosef}\ \bibnamefont
  {Zlochower}},\ }\bibfield  {title} {\enquote {\bibinfo {title} {{Remnant
  mass, spin, and recoil from spin aligned black-hole binaries}},}\ }\href
  {\doibase 10.1103/PhysRevD.90.104004} {\bibfield  {journal} {\bibinfo
  {journal} {Phys. Rev.}\ }\textbf {\bibinfo {volume} {D90}},\ \bibinfo {pages}
  {104004} (\bibinfo {year} {2014})},\ \Eprint {http://arxiv.org/abs/1406.7295}
  {arXiv:1406.7295 [gr-qc]} \BibitemShut {NoStop}%
\bibitem [{\citenamefont {Vishveshwara}(1970)}]{Vishveshwara:1970}%
  \BibitemOpen
  \bibfield  {author} {\bibinfo {author} {\bibfnamefont {C.~V.}\ \bibnamefont
  {Vishveshwara}},\ }\bibfield  {title} {\enquote {\bibinfo {title} {Stability
  of the schwarzschild metric},}\ }\href {\doibase 10.1103/PhysRevD.1.2870}
  {\bibfield  {journal} {\bibinfo  {journal} {Phys. Rev. D}\ }\textbf {\bibinfo
  {volume} {1}},\ \bibinfo {pages} {2870--2879} (\bibinfo {year}
  {1970})}\BibitemShut {NoStop}%
\bibitem [{\citenamefont {{Press}}(1971)}]{Press:1971ApJ}%
  \BibitemOpen
  \bibfield  {author} {\bibinfo {author} {\bibfnamefont {William~H.}\
  \bibnamefont {{Press}}},\ }\bibfield  {title} {\enquote {\bibinfo {title}
  {{Long Wave Trains of Gravitational Waves from a Vibrating Black Hole}},}\
  }\href {\doibase 10.1086/180849} {\bibfield  {journal} {\bibinfo  {journal}
  {Astrophysical Journal}\ }\textbf {\bibinfo {volume} {170}},\ \bibinfo
  {pages} {L105} (\bibinfo {year} {1971})}\BibitemShut {NoStop}%
\bibitem [{\citenamefont {{Teukolsky}}(1973)}]{Teukolsky:1973ApJ}%
  \BibitemOpen
  \bibfield  {author} {\bibinfo {author} {\bibfnamefont {Saul~A.}\ \bibnamefont
  {{Teukolsky}}},\ }\bibfield  {title} {\enquote {\bibinfo {title}
  {{Perturbations of a Rotating Black Hole. I. Fundamental Equations for
  Gravitational, Electromagnetic, and Neutrino-Field Perturbations}},}\ }\href
  {\doibase 10.1086/152444} {\bibfield  {journal} {\bibinfo  {journal}
  {Astrophysical Journal}\ }\textbf {\bibinfo {volume} {185}},\ \bibinfo
  {pages} {635--648} (\bibinfo {year} {1973})}\BibitemShut {NoStop}%
\bibitem [{\citenamefont {Chandrasekhar}\ and\ \citenamefont
  {Detweiler}(1975)}]{Chandrasekhar_Detweiler:1975}%
  \BibitemOpen
  \bibfield  {author} {\bibinfo {author} {\bibfnamefont {S.}~\bibnamefont
  {Chandrasekhar}}\ and\ \bibinfo {author} {\bibfnamefont {S.}~\bibnamefont
  {Detweiler}},\ }\bibfield  {title} {\enquote {\bibinfo {title} {The
  quasi-normal modes of the schwarzschild black hole},}\ }\href
  {http://www.jstor.org/stable/78902} {\bibfield  {journal} {\bibinfo
  {journal} {Proceedings of the Royal Society of London. Series A, Mathematical
  and Physical Sciences}\ }\textbf {\bibinfo {volume} {344}},\ \bibinfo {pages}
  {441--452} (\bibinfo {year} {1975})}\BibitemShut {NoStop}%
\bibitem [{\citenamefont {Gair}\ and\ \citenamefont
  {Moore}(2015)}]{Gair:2015nga}%
  \BibitemOpen
  \bibfield  {author} {\bibinfo {author} {\bibfnamefont {Jonathan~R.}\
  \bibnamefont {Gair}}\ and\ \bibinfo {author} {\bibfnamefont {Christopher~J.}\
  \bibnamefont {Moore}},\ }\bibfield  {title} {\enquote {\bibinfo {title}
  {{Quantifying and mitigating bias in inference on gravitational wave source
  populations}},}\ }\href {\doibase 10.1103/PhysRevD.91.124062} {\bibfield
  {journal} {\bibinfo  {journal} {Phys. Rev.}\ }\textbf {\bibinfo {volume}
  {D91}},\ \bibinfo {pages} {124062} (\bibinfo {year} {2015})},\ \Eprint
  {http://arxiv.org/abs/1504.02767} {arXiv:1504.02767 [gr-qc]} \BibitemShut
  {NoStop}%
\bibitem [{\citenamefont {Dreyer}\ \emph {et~al.}(2004)\citenamefont {Dreyer},
  \citenamefont {Kelly}, \citenamefont {Krishnan}, \citenamefont {Finn},
  \citenamefont {Garrison},\ and\ \citenamefont
  {Lopez-Aleman}}]{Dreyer:2003bv}%
  \BibitemOpen
  \bibfield  {author} {\bibinfo {author} {\bibfnamefont {Olaf}\ \bibnamefont
  {Dreyer}}, \bibinfo {author} {\bibfnamefont {Bernard~J.}\ \bibnamefont
  {Kelly}}, \bibinfo {author} {\bibfnamefont {Badri}\ \bibnamefont {Krishnan}},
  \bibinfo {author} {\bibfnamefont {Lee~Samuel}\ \bibnamefont {Finn}}, \bibinfo
  {author} {\bibfnamefont {David}\ \bibnamefont {Garrison}}, \ and\ \bibinfo
  {author} {\bibfnamefont {Ramon}\ \bibnamefont {Lopez-Aleman}},\ }\bibfield
  {title} {\enquote {\bibinfo {title} {{Black hole spectroscopy: Testing
  general relativity through gravitational wave observations}},}\ }\href
  {\doibase 10.1088/0264-9381/21/4/003} {\bibfield  {journal} {\bibinfo
  {journal} {Class. Quant. Grav.}\ }\textbf {\bibinfo {volume} {21}},\ \bibinfo
  {pages} {787--804} (\bibinfo {year} {2004})},\ \Eprint
  {http://arxiv.org/abs/gr-qc/0309007} {arXiv:gr-qc/0309007 [gr-qc]}
  \BibitemShut {NoStop}%
\bibitem [{\citenamefont {Gossan}\ \emph {et~al.}(2012)\citenamefont {Gossan},
  \citenamefont {Veitch},\ and\ \citenamefont {Sathyaprakash}}]{Gossan:2011ha}%
  \BibitemOpen
  \bibfield  {author} {\bibinfo {author} {\bibfnamefont {S.}~\bibnamefont
  {Gossan}}, \bibinfo {author} {\bibfnamefont {J.}~\bibnamefont {Veitch}}, \
  and\ \bibinfo {author} {\bibfnamefont {B.~S.}\ \bibnamefont
  {Sathyaprakash}},\ }\bibfield  {title} {\enquote {\bibinfo {title} {{Bayesian
  model selection for testing the no-hair theorem with black hole
  ringdowns}},}\ }\href {\doibase 10.1103/PhysRevD.85.124056} {\bibfield
  {journal} {\bibinfo  {journal} {Phys. Rev.}\ }\textbf {\bibinfo {volume}
  {D85}},\ \bibinfo {pages} {124056} (\bibinfo {year} {2012})},\ \Eprint
  {http://arxiv.org/abs/1111.5819} {arXiv:1111.5819 [gr-qc]} \BibitemShut
  {NoStop}%
\bibitem [{\citenamefont {Meidam}\ \emph {et~al.}(2014)\citenamefont {Meidam},
  \citenamefont {Agathos}, \citenamefont {Van Den~Broeck}, \citenamefont
  {Veitch},\ and\ \citenamefont {Sathyaprakash}}]{Meidam:2014jpa}%
  \BibitemOpen
  \bibfield  {author} {\bibinfo {author} {\bibfnamefont {J.}~\bibnamefont
  {Meidam}}, \bibinfo {author} {\bibfnamefont {M.}~\bibnamefont {Agathos}},
  \bibinfo {author} {\bibfnamefont {C.}~\bibnamefont {Van Den~Broeck}},
  \bibinfo {author} {\bibfnamefont {J.}~\bibnamefont {Veitch}}, \ and\ \bibinfo
  {author} {\bibfnamefont {B.~S.}\ \bibnamefont {Sathyaprakash}},\ }\bibfield
  {title} {\enquote {\bibinfo {title} {{Testing the no-hair theorem with black
  hole ringdowns using TIGER}},}\ }\href {\doibase 10.1103/PhysRevD.90.064009}
  {\bibfield  {journal} {\bibinfo  {journal} {Phys. Rev.}\ }\textbf {\bibinfo
  {volume} {D90}},\ \bibinfo {pages} {064009} (\bibinfo {year} {2014})},\
  \Eprint {http://arxiv.org/abs/1406.3201} {arXiv:1406.3201 [gr-qc]}
  \BibitemShut {NoStop}%
\bibitem [{\citenamefont {Gerosa}\ and\ \citenamefont
  {Moore}(2016)}]{Gerosa:2016vip}%
  \BibitemOpen
  \bibfield  {author} {\bibinfo {author} {\bibfnamefont {Davide}\ \bibnamefont
  {Gerosa}}\ and\ \bibinfo {author} {\bibfnamefont {Christopher~J.}\
  \bibnamefont {Moore}},\ }\bibfield  {title} {\enquote {\bibinfo {title}
  {{Black hole kicks as new gravitational wave observables}},}\ }\href
  {\doibase 10.1103/PhysRevLett.117.011101} {\bibfield  {journal} {\bibinfo
  {journal} {Phys. Rev. Lett.}\ }\textbf {\bibinfo {volume} {117}},\ \bibinfo
  {pages} {011101} (\bibinfo {year} {2016})},\ \Eprint
  {http://arxiv.org/abs/1606.04226} {arXiv:1606.04226 [gr-qc]} \BibitemShut
  {NoStop}%
\bibitem [{\citenamefont {Reitze}\ \emph {et~al.}(2019)\citenamefont {Reitze}
  \emph {et~al.}}]{Reitze:2019iox}%
  \BibitemOpen
  \bibfield  {author} {\bibinfo {author} {\bibfnamefont {David}\ \bibnamefont
  {Reitze}} \emph {et~al.},\ }\bibfield  {title} {\enquote {\bibinfo {title}
  {{Cosmic Explorer: The U.S. Contribution to Gravitational-Wave Astronomy
  beyond LIGO}},}\ }\href@noop {} {\bibfield  {journal} {\bibinfo  {journal}
  {Bull. Am. Astron. Soc.}\ }\textbf {\bibinfo {volume} {51}},\ \bibinfo
  {pages} {035} (\bibinfo {year} {2019})},\ \Eprint
  {http://arxiv.org/abs/1907.04833} {arXiv:1907.04833 [astro-ph.IM]}
  \BibitemShut {NoStop}%
\bibitem [{\citenamefont {Punturo}\ \emph
  {et~al.}(2010{\natexlab{a}})\citenamefont {Punturo} \emph
  {et~al.}}]{Punturo:2010zz}%
  \BibitemOpen
  \bibfield  {author} {\bibinfo {author} {\bibfnamefont {M.}~\bibnamefont
  {Punturo}} \emph {et~al.},\ }\bibfield  {title} {\enquote {\bibinfo {title}
  {{The Einstein Telescope: A third-generation gravitational wave
  observatory}},}\ }\bibfield  {booktitle} {\emph {\bibinfo {booktitle}
  {{Proceedings, 14th Workshop on Gravitational wave data analysis (GWDAW-14):
  Rome, Italy, January 26-29, 2010}}},\ }\href {\doibase
  10.1088/0264-9381/27/19/194002} {\bibfield  {journal} {\bibinfo  {journal}
  {Class. Quant. Grav.}\ }\textbf {\bibinfo {volume} {27}},\ \bibinfo {pages}
  {194002} (\bibinfo {year} {2010}{\natexlab{a}})}\BibitemShut {NoStop}%
\bibitem [{\citenamefont {Punturo}\ \emph
  {et~al.}(2010{\natexlab{b}})\citenamefont {Punturo} \emph
  {et~al.}}]{Punturo:2010science_reach}%
  \BibitemOpen
  \bibfield  {author} {\bibinfo {author} {\bibfnamefont {M}~\bibnamefont
  {Punturo}} \emph {et~al.},\ }\bibfield  {title} {\enquote {\bibinfo {title}
  {The third generation of gravitational wave observatories and their science
  reach},}\ }\href {\doibase 10.1088/0264-9381/27/8/084007} {\bibfield
  {journal} {\bibinfo  {journal} {Classical and Quantum Gravity}\ }\textbf
  {\bibinfo {volume} {27}},\ \bibinfo {pages} {084007} (\bibinfo {year}
  {2010}{\natexlab{b}})}\BibitemShut {NoStop}%
\bibitem [{\citenamefont {Abbott}\ \emph {et~al.}(2017)\citenamefont {Abbott}
  \emph {et~al.}}]{Evans:2016mbw}%
  \BibitemOpen
  \bibfield  {author} {\bibinfo {author} {\bibfnamefont {Benjamin~P}\
  \bibnamefont {Abbott}} \emph {et~al.} (\bibinfo {collaboration} {LIGO
  Scientific}),\ }\bibfield  {title} {\enquote {\bibinfo {title} {{Exploring
  the Sensitivity of Next Generation Gravitational Wave Detectors}},}\ }\href
  {\doibase 10.1088/1361-6382/aa51f4} {\bibfield  {journal} {\bibinfo
  {journal} {Class. Quant. Grav.}\ }\textbf {\bibinfo {volume} {34}},\ \bibinfo
  {pages} {044001} (\bibinfo {year} {2017})},\ \Eprint
  {http://arxiv.org/abs/1607.08697} {arXiv:1607.08697 [astro-ph.IM]}
  \BibitemShut {NoStop}%
\bibitem [{\citenamefont {Calderón~Bustillo}\ \emph
  {et~al.}(2018)\citenamefont {Calderón~Bustillo}, \citenamefont {Clark},
  \citenamefont {Laguna},\ and\ \citenamefont
  {Shoemaker}}]{CalderonBustillo:2018zuq}%
  \BibitemOpen
  \bibfield  {author} {\bibinfo {author} {\bibfnamefont {Juan}\ \bibnamefont
  {Calderón~Bustillo}}, \bibinfo {author} {\bibfnamefont {James~A.}\
  \bibnamefont {Clark}}, \bibinfo {author} {\bibfnamefont {Pablo}\ \bibnamefont
  {Laguna}}, \ and\ \bibinfo {author} {\bibfnamefont {Deirdre}\ \bibnamefont
  {Shoemaker}},\ }\bibfield  {title} {\enquote {\bibinfo {title} {{Tracking
  black hole kicks from gravitational wave observations}},}\ }\href {\doibase
  10.1103/PhysRevLett.121.191102} {\bibfield  {journal} {\bibinfo  {journal}
  {Phys. Rev. Lett.}\ }\textbf {\bibinfo {volume} {121}},\ \bibinfo {pages}
  {191102} (\bibinfo {year} {2018})},\ \Eprint
  {http://arxiv.org/abs/1806.11160} {arXiv:1806.11160 [gr-qc]} \BibitemShut
  {NoStop}%
\bibitem [{\citenamefont {Healy}\ \emph {et~al.}(2019)\citenamefont {Healy},
  \citenamefont {Lousto}, \citenamefont {Lange}, \citenamefont {O'Shaughnessy},
  \citenamefont {Zlochower},\ and\ \citenamefont {Campanelli}}]{Healy:2019}%
  \BibitemOpen
  \bibfield  {author} {\bibinfo {author} {\bibfnamefont {James}\ \bibnamefont
  {Healy}}, \bibinfo {author} {\bibfnamefont {Carlos~O.}\ \bibnamefont
  {Lousto}}, \bibinfo {author} {\bibfnamefont {Jacob}\ \bibnamefont {Lange}},
  \bibinfo {author} {\bibfnamefont {Richard}\ \bibnamefont {O'Shaughnessy}},
  \bibinfo {author} {\bibfnamefont {Yosef}\ \bibnamefont {Zlochower}}, \ and\
  \bibinfo {author} {\bibfnamefont {Manuela}\ \bibnamefont {Campanelli}},\
  }\href@noop {} {\enquote {\bibinfo {title} {The second {RIT} binary black
  hole simulations catalog and its application to gravitational waves parameter
  estimation},}\ } (\bibinfo {year} {2019}),\ \Eprint
  {http://arxiv.org/abs/1901.02553} {arXiv:1901.02553 [gr-qc]} \BibitemShut
  {NoStop}%
\bibitem [{\citenamefont {Torres-Orjuela}\ \emph {et~al.}(2020)\citenamefont
  {Torres-Orjuela}, \citenamefont {Chen},\ and\ \citenamefont
  {Amaro-Seoane}}]{Torres-Orjuela:2020cly}%
  \BibitemOpen
  \bibfield  {author} {\bibinfo {author} {\bibfnamefont {Alejandro}\
  \bibnamefont {Torres-Orjuela}}, \bibinfo {author} {\bibfnamefont {Xian}\
  \bibnamefont {Chen}}, \ and\ \bibinfo {author} {\bibfnamefont {Pau}\
  \bibnamefont {Amaro-Seoane}},\ }\bibfield  {title} {\enquote {\bibinfo
  {title} {{A phase shift of gravitational waves induced by aberration}},}\
  }\href@noop {} {\  (\bibinfo {year} {2020})},\ \Eprint
  {http://arxiv.org/abs/2001.00721} {arXiv:2001.00721 [astro-ph.HE]}
  \BibitemShut {NoStop}%
\bibitem [{\citenamefont {Abbott}\ \emph
  {et~al.}(2016{\natexlab{b}})\citenamefont {Abbott} \emph
  {et~al.}}]{Abbott:2016blz}%
  \BibitemOpen
  \bibfield  {author} {\bibinfo {author} {\bibfnamefont {B.~P.}\ \bibnamefont
  {Abbott}} \emph {et~al.} (\bibinfo {collaboration} {LIGO Scientific,
  Virgo}),\ }\bibfield  {title} {\enquote {\bibinfo {title} {{Observation of
  Gravitational Waves from a Binary Black Hole Merger}},}\ }\href {\doibase
  10.1103/PhysRevLett.116.061102} {\bibfield  {journal} {\bibinfo  {journal}
  {Phys. Rev. Lett.}\ }\textbf {\bibinfo {volume} {116}},\ \bibinfo {pages}
  {061102} (\bibinfo {year} {2016}{\natexlab{b}})},\ \Eprint
  {http://arxiv.org/abs/1602.03837} {arXiv:1602.03837 [gr-qc]} \BibitemShut
  {NoStop}%
\bibitem [{\citenamefont {Campanelli}\ \emph
  {et~al.}(2007{\natexlab{b}})\citenamefont {Campanelli}, \citenamefont
  {Lousto}, \citenamefont {Zlochower},\ and\ \citenamefont
  {Merritt}}]{Campanelli:2007ew}%
  \BibitemOpen
  \bibfield  {author} {\bibinfo {author} {\bibfnamefont {Manuela}\ \bibnamefont
  {Campanelli}}, \bibinfo {author} {\bibfnamefont {Carlos~O.}\ \bibnamefont
  {Lousto}}, \bibinfo {author} {\bibfnamefont {Yosef}\ \bibnamefont
  {Zlochower}}, \ and\ \bibinfo {author} {\bibfnamefont {David}\ \bibnamefont
  {Merritt}},\ }\bibfield  {title} {\enquote {\bibinfo {title} {{Large merger
  recoils and spin flips from generic black-hole binaries}},}\ }\href {\doibase
  10.1086/516712} {\bibfield  {journal} {\bibinfo  {journal} {Astrophys. J.}\
  }\textbf {\bibinfo {volume} {659}},\ \bibinfo {pages} {L5--L8} (\bibinfo
  {year} {2007}{\natexlab{b}})},\ \Eprint {http://arxiv.org/abs/gr-qc/0701164}
  {arXiv:gr-qc/0701164 [gr-qc]} \BibitemShut {NoStop}%
\bibitem [{\citenamefont {Gerosa}\ and\ \citenamefont
  {Kesden}(2016)}]{Gerosa:2016sys}%
  \BibitemOpen
  \bibfield  {author} {\bibinfo {author} {\bibfnamefont {Davide}\ \bibnamefont
  {Gerosa}}\ and\ \bibinfo {author} {\bibfnamefont {Michael}\ \bibnamefont
  {Kesden}},\ }\bibfield  {title} {\enquote {\bibinfo {title} {{PRECESSION:
  Dynamics of spinning black-hole binaries with python}},}\ }\href {\doibase
  10.1103/PhysRevD.93.124066} {\bibfield  {journal} {\bibinfo  {journal} {Phys.
  Rev.}\ }\textbf {\bibinfo {volume} {D93}},\ \bibinfo {pages} {124066}
  (\bibinfo {year} {2016})},\ \Eprint {http://arxiv.org/abs/1605.01067}
  {arXiv:1605.01067 [astro-ph.HE]} \BibitemShut {NoStop}%
\bibitem [{\citenamefont {{SXS Collaboration}}()}]{SXSCatalog}%
  \BibitemOpen
  \bibfield  {author} {\bibinfo {author} {\bibnamefont {{SXS Collaboration}}},\
  }\href@noop {} {\enquote {\bibinfo {title} {The {SXS} collaboration catalog
  of gravitational waveforms},}\ }\bibinfo {note}
  {\url{http://www.black-holes.org/waveforms}}\BibitemShut {NoStop}%
\bibitem [{\citenamefont {Mroue}\ \emph {et~al.}(2013)\citenamefont {Mroue}
  \emph {et~al.}}]{Mroue:2013xna}%
  \BibitemOpen
  \bibfield  {author} {\bibinfo {author} {\bibfnamefont {Abdul~H.}\
  \bibnamefont {Mroue}} \emph {et~al.},\ }\bibfield  {title} {\enquote
  {\bibinfo {title} {{Catalog of 174 Binary Black Hole Simulations for
  Gravitational Wave Astronomy}},}\ }\href {\doibase
  10.1103/PhysRevLett.111.241104} {\bibfield  {journal} {\bibinfo  {journal}
  {Phys. Rev. Lett.}\ }\textbf {\bibinfo {volume} {111}},\ \bibinfo {pages}
  {241104} (\bibinfo {year} {2013})},\ \Eprint {http://arxiv.org/abs/1304.6077}
  {arXiv:1304.6077 [gr-qc]} \BibitemShut {NoStop}%
\bibitem [{\citenamefont {Boyle}\ \emph {et~al.}(2019)\citenamefont {Boyle}
  \emph {et~al.}}]{Boyle:2019kee}%
  \BibitemOpen
  \bibfield  {author} {\bibinfo {author} {\bibfnamefont {Michael}\ \bibnamefont
  {Boyle}} \emph {et~al.},\ }\bibfield  {title} {\enquote {\bibinfo {title}
  {{The SXS Collaboration catalog of binary black hole simulations}},}\ }\href
  {\doibase 10.1088/1361-6382/ab34e2} {\bibfield  {journal} {\bibinfo
  {journal} {Class. Quant. Grav.}\ }\textbf {\bibinfo {volume} {36}},\ \bibinfo
  {pages} {195006} (\bibinfo {year} {2019})},\ \Eprint
  {http://arxiv.org/abs/1904.04831} {arXiv:1904.04831 [gr-qc]} \BibitemShut
  {NoStop}%
\bibitem [{\citenamefont {Collaboration}\ and\ \citenamefont
  {Collaboration}()}]{GW_open_science_center}%
  \BibitemOpen
  \bibfield  {author} {\bibinfo {author} {\bibfnamefont {LIGO~Scientific}\
  \bibnamefont {Collaboration}}\ and\ \bibinfo {author} {\bibfnamefont {Virgo}\
  \bibnamefont {Collaboration}},\ }\bibfield  {title} {\enquote {\bibinfo
  {title} {{Gravitational Wave Open Science Center}},}\ }\href@noop {} {\
  }\bibinfo {note} {\url{https://www.gw-openscience.org}}\BibitemShut {NoStop}%
\bibitem [{\citenamefont {Kullback}\ and\ \citenamefont
  {Leibler}(1951)}]{Kullback_Leibler_divergence}%
  \BibitemOpen
  \bibfield  {author} {\bibinfo {author} {\bibfnamefont {S.}~\bibnamefont
  {Kullback}}\ and\ \bibinfo {author} {\bibfnamefont {R.~A.}\ \bibnamefont
  {Leibler}},\ }\bibfield  {title} {\enquote {\bibinfo {title} {On information
  and sufficiency},}\ }\href {http://www.jstor.org/stable/2236703} {\bibfield
  {journal} {\bibinfo  {journal} {The Annals of Mathematical Statistics}\
  }\textbf {\bibinfo {volume} {22}},\ \bibinfo {pages} {79--86} (\bibinfo
  {year} {1951})}\BibitemShut {NoStop}%
\bibitem [{\citenamefont {Schmidt}\ \emph {et~al.}(2015)\citenamefont
  {Schmidt}, \citenamefont {Ohme},\ and\ \citenamefont
  {Hannam}}]{Schmidt:2014iyl}%
  \BibitemOpen
  \bibfield  {author} {\bibinfo {author} {\bibfnamefont {P.}~\bibnamefont
  {Schmidt}}, \bibinfo {author} {\bibfnamefont {F.}~\bibnamefont {Ohme}}, \
  and\ \bibinfo {author} {\bibfnamefont {M.}~\bibnamefont {Hannam}},\
  }\bibfield  {title} {\enquote {\bibinfo {title} {{Towards models of
  gravitational waveforms from generic binaries II: Modelling precession
  effects with a single effective precession parameter}},}\ }\href {\doibase
  10.1103/PhysRevD.91.024043} {\bibfield  {journal} {\bibinfo  {journal} {Phys.
  Rev. D}\ }\textbf {\bibinfo {volume} {91}},\ \bibinfo {pages} {024043}
  (\bibinfo {year} {2015})},\ \Eprint {http://arxiv.org/abs/1408.1810}
  {arXiv:1408.1810 [gr-qc]} \BibitemShut {NoStop}%
\bibitem [{\citenamefont {Bruegmann}\ \emph {et~al.}(2008)\citenamefont
  {Bruegmann}, \citenamefont {Gonzalez}, \citenamefont {Hannam}, \citenamefont
  {Husa},\ and\ \citenamefont {Sperhake}}]{Brugmann:2007zj}%
  \BibitemOpen
  \bibfield  {author} {\bibinfo {author} {\bibfnamefont {Bernd}\ \bibnamefont
  {Bruegmann}}, \bibinfo {author} {\bibfnamefont {Jose~A.}\ \bibnamefont
  {Gonzalez}}, \bibinfo {author} {\bibfnamefont {Mark}\ \bibnamefont {Hannam}},
  \bibinfo {author} {\bibfnamefont {Sascha}\ \bibnamefont {Husa}}, \ and\
  \bibinfo {author} {\bibfnamefont {Ulrich}\ \bibnamefont {Sperhake}},\
  }\bibfield  {title} {\enquote {\bibinfo {title} {{Exploring black hole
  superkicks}},}\ }\href {\doibase 10.1103/PhysRevD.77.124047} {\bibfield
  {journal} {\bibinfo  {journal} {Phys. Rev.}\ }\textbf {\bibinfo {volume}
  {D77}},\ \bibinfo {pages} {124047} (\bibinfo {year} {2008})},\ \Eprint
  {http://arxiv.org/abs/0707.0135} {arXiv:0707.0135 [gr-qc]} \BibitemShut
  {NoStop}%
\bibitem [{\citenamefont {Zlochower}\ and\ \citenamefont
  {Lousto}(2015)}]{Zlochower:2015wga}%
  \BibitemOpen
  \bibfield  {author} {\bibinfo {author} {\bibfnamefont {Yosef}\ \bibnamefont
  {Zlochower}}\ and\ \bibinfo {author} {\bibfnamefont {Carlos~O.}\ \bibnamefont
  {Lousto}},\ }\bibfield  {title} {\enquote {\bibinfo {title} {{Modeling the
  remnant mass, spin, and recoil from unequal-mass, precessing black-hole
  binaries: The Intermediate Mass Ratio Regime}},}\ }\href {\doibase
  10.1103/PhysRevD.92.024022} {\bibfield  {journal} {\bibinfo  {journal} {Phys.
  Rev.}\ }\textbf {\bibinfo {volume} {D92}},\ \bibinfo {pages} {024022}
  (\bibinfo {year} {2015})},\ \bibinfo {note} {[Erratum: Phys.
  Rev.D94,no.2,029901(2016)]},\ \Eprint {http://arxiv.org/abs/1503.07536}
  {arXiv:1503.07536 [gr-qc]} \BibitemShut {NoStop}%
\bibitem [{\citenamefont {Gerosa}\ \emph {et~al.}(2018)\citenamefont {Gerosa},
  \citenamefont {Hébert},\ and\ \citenamefont {Stein}}]{Gerosa:2018qay}%
  \BibitemOpen
  \bibfield  {author} {\bibinfo {author} {\bibfnamefont {Davide}\ \bibnamefont
  {Gerosa}}, \bibinfo {author} {\bibfnamefont {François}\ \bibnamefont
  {Hébert}}, \ and\ \bibinfo {author} {\bibfnamefont {Leo~C.}\ \bibnamefont
  {Stein}},\ }\bibfield  {title} {\enquote {\bibinfo {title} {{Black-hole kicks
  from numerical-relativity surrogate models}},}\ }\href {\doibase
  10.1103/PhysRevD.97.104049} {\bibfield  {journal} {\bibinfo  {journal} {Phys.
  Rev.}\ }\textbf {\bibinfo {volume} {D97}},\ \bibinfo {pages} {104049}
  (\bibinfo {year} {2018})},\ \Eprint {http://arxiv.org/abs/1802.04276}
  {arXiv:1802.04276 [gr-qc]} \BibitemShut {NoStop}%
\bibitem [{\citenamefont {Varma}\ \emph
  {et~al.}(2019{\natexlab{c}})\citenamefont {Varma}, \citenamefont {Stein},\
  and\ \citenamefont {Gerosa}}]{Varma:2018rcg}%
  \BibitemOpen
  \bibfield  {author} {\bibinfo {author} {\bibfnamefont {Vijay}\ \bibnamefont
  {Varma}}, \bibinfo {author} {\bibfnamefont {Leo~C.}\ \bibnamefont {Stein}}, \
  and\ \bibinfo {author} {\bibfnamefont {Davide}\ \bibnamefont {Gerosa}},\
  }\bibfield  {title} {\enquote {\bibinfo {title} {{The binary black hole
  explorer: on-the-fly visualizations of precessing binary black holes}},}\
  }\href {\doibase 10.1088/1361-6382/ab0ee9} {\bibfield  {journal} {\bibinfo
  {journal} {Class. Quant. Grav.}\ }\textbf {\bibinfo {volume} {36}},\ \bibinfo
  {pages} {095007} (\bibinfo {year} {2019}{\natexlab{c}})},\ \Eprint
  {http://arxiv.org/abs/1811.06552} {arXiv:1811.06552 [astro-ph.HE]}
  \BibitemShut {NoStop}%
\bibitem [{kic({\natexlab{b}})}]{kickpaperminkickfootnote}%
  \BibitemOpen
  \href@noop {} {}\bibinfo {howpublished} {The overall minimum should be zero,
  but a 35 km/s limit arises from numerical noise in the simulations on which
  \NRSurRemnant is trained. In spite of this, \NRSurRemnant is more accurate
  than alternate kick models by an order-of-magnitude ~\cite{Varma:2019csw}.}
  ({\natexlab{b}})\BibitemShut {NoStop}%
\bibitem [{\citenamefont {Berti}\ \emph {et~al.}(2012)\citenamefont {Berti},
  \citenamefont {Kesden},\ and\ \citenamefont {Sperhake}}]{Berti:2012zp}%
  \BibitemOpen
  \bibfield  {author} {\bibinfo {author} {\bibfnamefont {Emanuele}\
  \bibnamefont {Berti}}, \bibinfo {author} {\bibfnamefont {Michael}\
  \bibnamefont {Kesden}}, \ and\ \bibinfo {author} {\bibfnamefont {Ulrich}\
  \bibnamefont {Sperhake}},\ }\bibfield  {title} {\enquote {\bibinfo {title}
  {{Effects of post-Newtonian Spin Alignment on the Distribution of Black-Hole
  Recoils}},}\ }\href {\doibase 10.1103/PhysRevD.85.124049} {\bibfield
  {journal} {\bibinfo  {journal} {Phys. Rev.}\ }\textbf {\bibinfo {volume}
  {D85}},\ \bibinfo {pages} {124049} (\bibinfo {year} {2012})},\ \Eprint
  {http://arxiv.org/abs/1203.2920} {arXiv:1203.2920 [astro-ph.HE]} \BibitemShut
  {NoStop}%
\bibitem [{\citenamefont {Antonini}\ and\ \citenamefont
  {Rasio}(2016)}]{Antonini:2016gqe}%
  \BibitemOpen
  \bibfield  {author} {\bibinfo {author} {\bibfnamefont {Fabio}\ \bibnamefont
  {Antonini}}\ and\ \bibinfo {author} {\bibfnamefont {Frederic~A.}\
  \bibnamefont {Rasio}},\ }\bibfield  {title} {\enquote {\bibinfo {title}
  {{Merging black hole binaries in galactic nuclei: implications for
  advanced-LIGO detections}},}\ }\href {\doibase 10.3847/0004-637X/831/2/187}
  {\bibfield  {journal} {\bibinfo  {journal} {Astrophys. J.}\ }\textbf
  {\bibinfo {volume} {831}},\ \bibinfo {pages} {187} (\bibinfo {year}
  {2016})},\ \Eprint {http://arxiv.org/abs/1606.04889} {arXiv:1606.04889
  [astro-ph.HE]} \BibitemShut {NoStop}%
\end{thebibliography}%

\clearpage
\section*{Supplemental materials}
\label{supp_mat}
\renewcommand{\sinekicksfignum}{\ref{fig:superkick}\xspace}
\renewcommand{\violinfignum}{\ref{fig:random_cases}\xspace}
\renewcommand{\dopplerfignum}{\ref{fig:doppler_mass}\xspace}
\setcounter{equation}{0}
\setcounter{figure}{0}
\setcounter{table}{0}
\renewcommand{\theequation}{S\arabic{equation}}
\renewcommand{\thefigure}{S\arabic{figure}}

\subsection{Implications for tests of general relativity}
At leading order, the kick's effect on the GW signal can be described as a
Doppler shift of the GW frequency $f$~\cite{Gerosa:2016vip}. Because general
relativity lacks any intrinsic length scales, a uniform increase in signal
frequency is completely degenerate with a decrease in total mass $M$, and
vice versa. Thus, if not explicitly accounted for, a frequency shift due to a
kick will bias mass measurements. This is analogous to the effect of the
cosmological redshift $z$ on the GWs: GW measurements only measure the
combination $M (1+z)$ known as the detector-frame mass, and the source-frame
mass is only inferred after assuming a cosmology~\cite{Krolak1987}. One
important difference between the cosmological and kick redshifts is that, in
the latter, the Doppler shift occurs only when the kick is imparted, mostly
near the merger~\cite{Gonzalez:2006md, Lousto:2007db, Lousto:2012su,
Lousto:2012gt}. Therefore the Doppler shift only affects the merger and
ringdown part of the signal, while a cosmological redshift rescales the GW
signal as a whole.

The amount of Doppler shift depends on the projection of the kick velocity
along the line of sight. At leading order, the Doppler-shifted remnant mass is
given by~\cite{Gerosa:2016vip}:
\begin{equation}
m^{\text{DS}}_f = m_f\,(1 + \bv_f \cdot \hat{\bm{n}}/c),
\label{eq:doppler_mass}
\end{equation}
where $c$ is the speed of light and $\hat{\bm{n}}$ is the unit vector pointing
from the observer to the source.  From our inference setup, we obtain posterior
distributions for the component parameters $\bL = \{m_1, m_2, \bchi_1,
\bchi_2\}$, as well as the line-of-sight parameters ($\iota$,
$\phi_{\mathrm{ref}}$). Our method to measure the kick recovers the full kick
vector $\bv_f$ given $\bL$. For each posterior sample, we then project the kick
along the line of sight to obtain the Doppler-shifted remnant mass.

The Doppler shift due to the kick velocity can play an important role in tests
of general relativity using the ringdown signal~\cite{Dreyer:2003bv,
Gossan:2011ha, Meidam:2014jpa, TheLIGOScientific:2016src,
LIGOScientific:2019fpa, Ghosh:2017gfp, Brito:2018rfr, Carullo:2019flw,
Isi:2019aib, Giesler:2019uxc}. In some of these tests, the remnant mass and
spin are measured from different portions of the signal and compared against
each other to check for consistency.  In one version of the test, the full
inspiral-merger-ringdown signal is first analyzed using a waveform model and
posterior distributions are obtained for $\bL$.  These are then passed to
fitting formulae (e.g. \cite{Barausse:2012qz, Hofmann:2016yih,
Jimenez-Forteza:2016oae, Healy:2016lce, Healy:2014yta}) for the remnant mass
and spin to obtain posterior distributions for these quantities. Finally,
considering only the ringdown signal and varying the quasi-normal-mode
frequencies~\cite{Vishveshwara:1970, Press:1971ApJ, Teukolsky:1973ApJ,
Chandrasekhar_Detweiler:1975}, the remnant mass and spin are independently
measured \cite{Dreyer:2003bv,
Gossan:2011ha,TheLIGOScientific:2016src,Brito:2018rfr, Carullo:2019flw,
Isi:2019aib}.

In the first case the inferred remnant mass is not sensitive to the Doppler
shift as traditional fitting formulas for the remnant mass do not account for
this. Apart from modeling errors, this is equivalent to measuring the remnant
mass from the apparent horizon of the remnant black hole in an NR
simulation~\cite{Boyle:2019kee}. In the second case, however, the observed
ringdown frequencies would be Doppler-shifted and the inferred remnant mass
would be the Doppler-shifted value. For large Doppler shifts, these two
measurements of the remnant mass would be inconsistent, mimicking a deviation
from general relativity.

Fig.~\dopplerfignum in the main document shows the remnant mass posterior
distribution before and after the Doppler shift for the superkick configuration
of Fig.~\sinekicksfignum. The \NRSurRemnant model is used to predict the kick
vector and the remnant mass before the Doppler shift, while
Eq.~(\ref{eq:doppler_mass}) is used to predict the Doppler-shifted remnant
mass. The two mass distributions are visibly different in Fig.~\dopplerfignum,
therefore this will be important to account for in ringdown tests of general
relativity. However, this is a fairly fine-tuned source configuration with a
large kick velocity. This effect is expected to become important when the
measurement precision for the remnant mass is comparable or smaller than the
Doppler shift, $\delta m_f/m_f \lesssim |\bv_f \cdot \hat{\bm{n}}/c|$.  Unless
signals with kick magnitudes of order $1000$ km/s are detected, we expect that
this effect will only be important for third-generation GW detectors. In any
case, our method can already be used to account for this effect in tests of
general relativity.

\subsection{Probability-Probability plots}
\label{sec:pp_plot}

To demonstrate the robustness of our Bayesian inference infrastructure using
the \NRSur and \NRSurRemnant models, we produce a probability-probability
(P-P) plot for the kick velocity, from a set of 87 simulated binary BH
injections into design-sensitivity Gaussian noise for a LIGO
Hanford-Livingston-Virgo detector network. (See, e.g.~Ref.~\cite{Gair:2015nga}
for an example of P-P plots in the context of GW data analysis.) For each
injection, we run the \textsc{LALInference} parameter estimation
package~\cite{Veitch:2014wba} to obtain posteriors for the binary parameters,
like the masses and spins ($\bL$). From those, we then derive posteriors on the
kick parameters using the \NRSurRemnant surrogate. The P-P plot shows the
fraction of events for which the posterior for a given parameter recovers the
true value at a particular credible interval, as a function of the credible
interval. If the posteriors are sampled successfully, the P-P plot should be
diagonal---meaning that the true value is recovered within the $x\%$-credible
interval $x\%$ of the time, consistent with statistical error.

\begin{figure}[thb]
\includegraphics[width=0.44\textwidth]{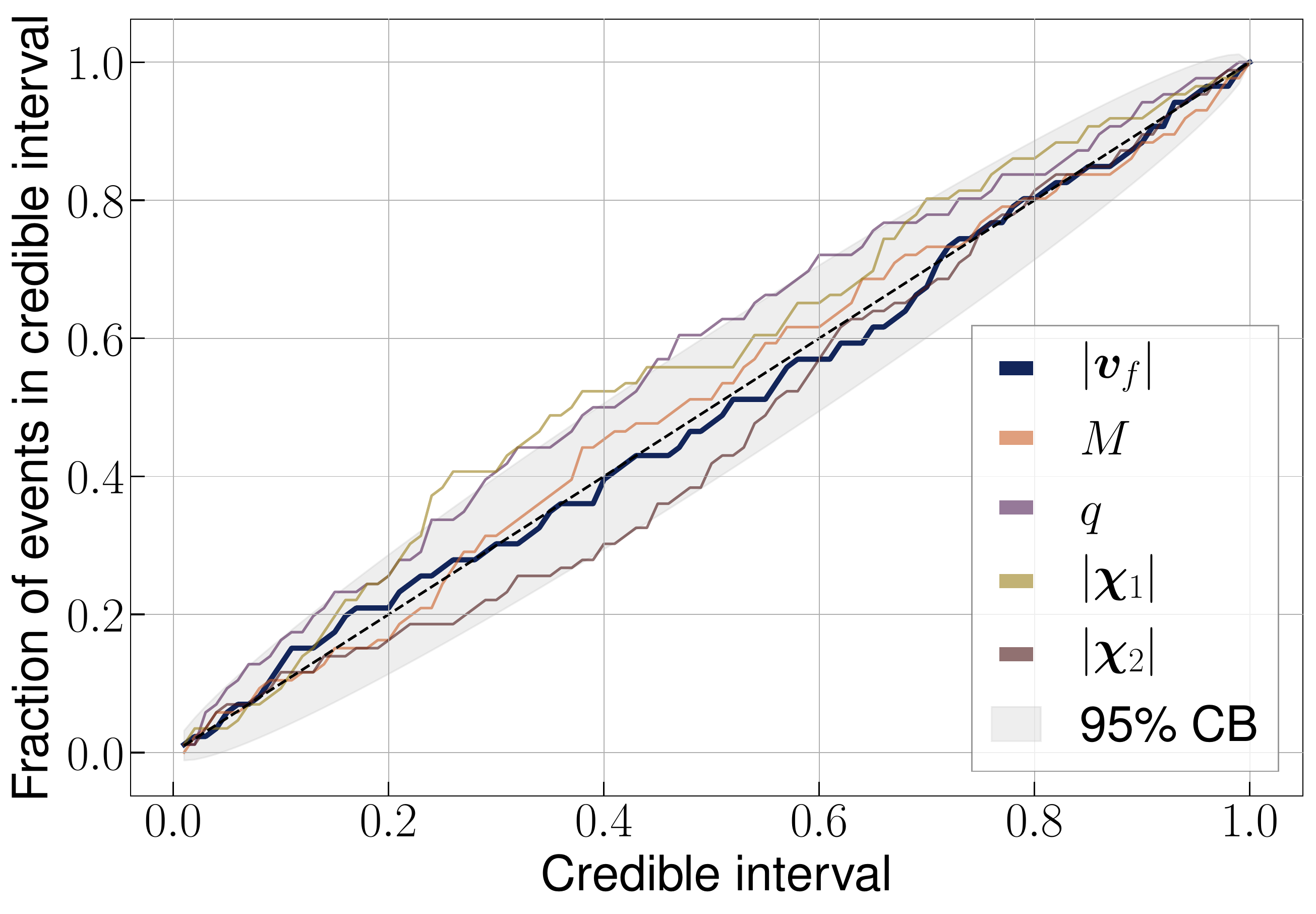}
\caption{P-P plot for 87 binary black hole injections into design sensitivity
    Gaussian noise recovered with the \NRSur and \NRSurRemnant models for
    the kick magnitude, as well as the total mass, mass ratio, and spin
    magnitudes. The diagonal is shown in the black dashed line along with the
    95\% confidence band (CB) in the shaded gray.
}
\label{fig:pp_plot}
\end{figure}

We draw the 87 injections from a distribution uniform in component masses
$m_1$, $m_2$ between 18 and $110~M_{\odot}$, but restricted to a mass ratio of
$q=m_1/m_2\leq 3$ and total mass $M \geq 72 \,M_{\odot}$. The spin magnitudes
are drawn uniformly between $ 0 \leq |\bchi_1|, |\bchi_2| \leq 0.8$, and the
directions are distributed isotropically on a sphere. The luminosity distances
are picked with a density proportional to their square (that is, uniform in
volume) out to 5 Gpc, and the inclination angle is drawn from a
uniform-in-cosine distribution. The location of the source in the sky is drawn
isotropically, as is its polarization angle. The same distributions are used as
the priors during the parameter estimation step. 

The P-P plot for the 87 simulated injections is shown in
Fig.~\ref{fig:pp_plot}. We display the distributions for the kick magnitude
($|\bv_f|$), total mass ($M$), mass ratio ($q$), and component spin parameters
$|\bchi_1|, |\bchi_2|$, which all lie largely within the 95\% confidence band
around the diagonal (shaded in gray). The p-value for the probability that the
fraction of events within a given credible interval for the kick magnitude is
drawn from uniform distribution between 0 and 1, as expected for a diagonal
P-P plot, is 98.6\%. This demonstrates that the kick posteriors generated with
\NRSurRemnant, in combination with the \textsc{LALInference} sampler, are
statistically robust and behave as expected in simulated Gaussian noise.
Deviations between the true value and the peak of the recovered posterior, such
as those seen in Fig.~\violinfignum, are consistent with statistical
fluctuations.\\

\subsection{Prior distribution for component parameters}

Analyses presented in the main text, for both injections and real data, use similar priors to those described in Sec.~\ref{sec:pp_plot}.
This choice follows standard conventions for LIGO-Virgo analyses (e.g., see Appendix C in \cite{LIGOScientific:2018mvr}).
We vary the specific mass ranges allowed to ensure the posterior always has full support within the prior.
For injections, the prior was uniform in component masses $m_1$, $m_2$ between 10 and $120~M_{\odot}$, but restricted to mass ratios $q=m_1/m_2\leq 4$ and total masses $M \geq 60 \,M_{\odot}$.
The spin magnitudes are drawn uniformly between $ 0 \leq |\bchi_1|, |\bchi_2| \leq 0.99$, and the directions are distributed isotropically on a sphere.
The priors on the extrinsic parameters are the same as in Sec.~\ref{sec:pp_plot}.

\end{document}